\documentclass[11pt,a4paper]{article}

\usepackage{jheppub}
\usepackage{macros}
\preprint{}

\title{On knots, complements, and \boldmath $6j$-symbols}
\author{Hao Ellery Wang,}
\author{Yuanzhe Jack Yang,}
\author{Hao Derrick Zhang,}
\author{and Satoshi Nawata}
\affiliation{Department of Physics and Center for Field Theory and Particle Physics, Fudan University, 220, Handan Road, 200433 Shanghai, China}
\date{}

\emailAdd{yukawahaow@gmail.com, jack\_yyz@outlook.com, haozhangphys@gmail.com, snawata@gmail.com}

\abstract{This paper investigates the relation between colored HOMFLY-PT and Kauffman homology, $\SO(N)$ quantum $6j$-symbols and  $(a,t)$-deformed $F_K$. First, we present a simple rule of grading change which allows us to obtain the $[r]$-colored quadruply-graded Kauffman homology from the $[r^2]$-colored quadruply-graded HOMFLY-PT homology for thin knots. This rule stems from the isomorphism of the representations $(\mathfrak{so}_6,[r]) \cong (\mathfrak{sl}_4,[r^2])$.
Also, we find the relationship among $A$-polynomials of $\SO$ and $\SU$-type coming from a differential on Kauffman homology.
Second, we  put forward a closed-form expression of $\SO(N)(N\geq 4)$ quantum $6j$-symbols for symmetric representations, and calculate the corresponding $\SO(N)$ fusion matrices for the cases when representations $R = $ {\tiny \yng(1)}, {\tiny \yng(2)}. Third, we conjecture closed-form expressions of $(a,t)$-deformed $F_K$ for the complements of double twist knots with positive braids. Using the conjectural expressions, we derive $t$-deformed ADO polynomials. }


\begin{document}

\allowdisplaybreaks
\maketitle

\Yboxdim5pt

\section{Introduction}
Witten's celebrated paper \cite{Witten:1988hf} has shown that the Chern-Simons theory provides a natural framework for quantum invariants of 3-manifolds and knots. The Chern-Simons path integral on a 3-manifold $M_3$ with the action
$$
S=\frac{k}{4\pi}\int_{M_3} \text{Tr}\left(A\wedge dA + \frac23 A\wedge A\wedge A\right)~,
$$
is a quantum invariant of $M_3$, known as the Witten-Reshetikhin-Turaev (WRT) invariant \cite{Witten:1988hf,Reshetikhin:1991tc}. If a Wilson loop is included along a knot in $S^3$, the expectation value of the Wilson loop is a quantum invariant of the knot:
$$
\overline{J}^G_R(K;q):=\left<W_R(K)\right>=\frac{\int \left[ \mathcal{D}A\right]e^{iS}W_R(K)}{\int\left[\mathcal{D}A\right]e^{iS}}~,
$$
where the parameter $q$ is expressed by the level $k$ and the dual Coxeter number $h^{\lor}$ of the gauge group $G$ as
\be\label{convention}
q=\exp\left(\frac{\pi i}{k+h^\lor}\right)~.
\ee
Since the Chern-Simons functional integral on a three-manifold with boundary is an element of the Hilbert space on the boundary which is isomorphic to the space of WZNW conformal blocks, quantum knot invariants could be constructed by using braiding and fusion operations on WZNW conformal blocks \cite{ Moore:1989vd,Kaul:1991np, Kaul:1992rs,Kaul:1993hb, Ramadevi:1992np}. When $G=\SU(N)$, it corresponds to a two-variable colored HOMFLY-PT polynomial $P_R(K;a,q)$, where $R$ is a representation of $\SU(N)$ and $a=q^N$. When $G=\SO(N)$, it corresponds to colored Kauffman polynomial $F_R(K;a,q)$, where $R$ is a representation of $\SO(N)$ and $a=q^{N-1}$.\footnote{
In this paper, a representation specified by a Young diagram $(r_1 \ge r_2 \ge \cdots)$ is denoted by $R=[r_1, r_2, \cdots]$. In particular, we write $[r^s]$ for an $r \times s$ rectangular Young diagram.} The quantum knot invariants  have a beautiful property that they are polynomials of $q$ with integer coefficients. This property led to an important development in knot theory proposed by Khovanov \cite{Khovanov:1999qla}, i.e. the categorifications of quantum knot invariants. Khovanov constructed  a bi-graded homology which itself is a knot invariant, and its $q$-graded Euler characteristic is the Jones polynomial $J(K;q)$ of a knot $K$.\footnote{We define the normalized quantum invariant
$$
J^G_R(K;q)=\frac{\overline{J}^G_R(K;q)}{\overline{J}^G_R({\raisebox{-.1cm}{\includegraphics[width=.4cm]{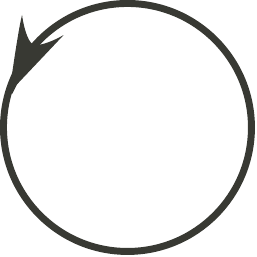}}};q)}~.
$$ In this paper, we focus only on knot invariants normalized by the unknot so that HOMFLY-PT $P_R(K;a,q)$ and Kauffman $F_R(K;a,q)$ polynomials are also normalized in a similar manner.}

In \cite{khovanov2008matrix},  M. Khovanov and L. Rozansky constructed a triply-graded homology of a knot whose graded Euler characteristic is the HOMFLY-PT polynomial. And in  \cite{Dunfield:2005si}, it was conjectured that the HOMFLY-PT homology is endowed with structural properties. Similarly, the existence and structural properties of Kauffman homology were conjectured in \cite{Gukov:2005qp} although the rigorous definition of a triply-graded homology theory categorifying the Kauffman polynomials has not been given yet.

This line of studying knot homology has been pursued in the colored cases. In \cite{Gukov:2011ry, Gorsky:2013jxa}, quadruply-graded colored HOMFLY-PT homology was proposed and various structures and symmetries were uncovered.
In \cite{Nawata:2013mzx}, it was further conjectured that there exist the quadruply-graded colored Kauffman homologies with rich structural properties.
The first part of this paper investigates the relationship between the colored HOMFLY-PT homology and Kauffman homology. In \S\ref{section:change-of-variable}, we show a rule of changing variables that transforms the Poincar\'e polynomial of $[r^2]$-colored  HOMFLY-PT homology into that of $[r]$-colored  Kauffman homology for thin knots.

As another important development,
the volume conjecture \cite{Kashaev:1996kc,murakami2001colored} gives a remarkable relationship between quantum invariants of a knot $K$ and classical geometry of the knot complement $S^3\backslash K$. It states that the large color asymptotic behavior of colored Jones polynomial $J_{[r]}(K;q)$ of a hyperbolic knot provides the hyperbolic volume of its knot complement. This was further generalized in \cite{Gukov:2003na,Gukov:2008jlx}, which gives the relation between the large color limit of colored Jones polynomials and the $A$-polynomial for the knot complement.
The zero locus of the $A$-polynomial $A(K;x,y)$ of a knot $K$ determines the character variety of $\SL(2,\mathbb{C})$-representation of the knot group \cite{Cooper:1994pca} so that the generalized volume conjecture paves a way to complex Chern-Simons theory.
There are further generalizations of the volume conjecture to incorporate HOMFLY-PT and Kauffman polynomials colored by symmetric representations and their categorifications \cite{Fuji:2012pm,Aganagic:2012jb,Fuji:2012nx,Nawata:2013mzx}, leading to a notion of super-$A$-polynomials. Since there is a differential that relates colored Kauffman homology to HOMFLY-PT homology, we investigate the influence of the differential on super-$A$-polynomials in \S\ref{section:super-a-polynomial}. This brings about Conjecture \ref{A-SO-SU} that provides an intriguing relationship among $A$-polynomials of $\SU$ and $\SO$-type.

As briefly mentioned, quantum invariants can be evaluated with braiding and fusion matrices in the WZNW models. When $q$ is a root of unity \eqref{convention}, the fusion matrices of the WZNW models are equivalent to quantum $6j$-symbols of the corresponding quantum groups (up to normalizations \eqref{6j}) \cite{Moore:1988qv, Moore:1989vd, kazhdan1993tensor, kazhdan1993tensor2, kazhdan1994tensor, kazhdan1994tensor2}. The $6j$-symbols were introduced by Wigner \cite{1965Quantum, wigner1993matrices} and Racah \cite{Racah:1942gsc} in the study of angular momenta, and they are beautifully expressed by a hypergeometric series ${}_4F_3$.  In the case of $\mathfrak{sl}_2$, the explicit expression for quantum $6j$-symbols was obtained in \cite{Kirillov:1991ec}, which is a natural quantization with a $q$-hypergeometric series ${}_4\varphi_3$. Nonetheless, it is notoriously involved to obtain explicit expressions for $6j$-symbols from the definition with the Clebsch-Gordan coefficients ($3j$-symbols), in particular, for higher ranks \cite{Gu:2014nva}. However, the information of quantum $6j$-symbols with all symmetric representations was extracted from quantum knot invariants for $U_q(\mathfrak{sl}_N)$ \cite{Nawata:2013ooa}, and it was formulated in terms of the $q$-hypergeometric series ${}_4\varphi_3$ in \cite{Alekseev:2019nzw}.
 In \S\ref{section:6j-symbols}, we study the $\SO(N)$ quantum $6j$-symbols with symmetric representations along this line.
The formulae for the classical $\SO(N)$ $6j$-symbols with all representations symmetric were obtained by Ali\v{s}auskas \cite{alisauskas1987some, Alisauskas_2002}. We quantize these formulae and compute $\SO(N)$ fusion matrices of the WZNW models for the case of  $R = $ { \yng(1)}, { \yng(2)}. We check that the fusion matrices reproduce the colored Kauffman polynomials of non-torus knots. This gives a piece of evidence that the natural quantization of the Ali\v{s}auskas' formulae will be valid for the $\SO(N)$ quantum $6j$-symbols.

The developments of knot invariants and volume conjectures are inseparably bound up with the study of invariants of 3-manifolds. Chern-Simons quantum invariants of 3-manifolds can be computed by a surgery formula of quantum invariants of links in $S^3$ \cite{Reshetikhin:1991tc} when $q$ is a root of unity \eqref{convention}. In the recent study of the 3d/3d correspondence, an analytic continuation $\widehat Z$ of the WRT invariant has been proposed as a BPS $q$-series \cite{Gukov:2016gkn, Gukov:2017kmk}. This invariant can be thought of as a generating series of BPS states in a 3d $\cN=2$ theory $T[M_3]$. Also it can be regarded as a partition function of complex Chern-Simons theory on $M_3$ defined on a unit disk $|q|<1$ whose radial limit $q \searrow e^{\pi i/(k+h^\vee)}$ becomes the WRT invariant. Its generalization to a knot complement, denoted as $F_K(x,q)=\widehat Z(S^3\backslash K)$, is given in \cite{Gukov:2019mnk} to make surgeries and TQFT method work.
It is observed that the asymptotic expansion of $F_K$ agrees with the Melvin-Morton-Rozansky expansion \cite{melvin1995,Stavros1996,Rozansky:1994qe,Rozansky98theuniversal} of
the colored Jones polynomials. Moreover, $F_K$ is conjectured to be annihilated by the quantum $A$-polynomial \cite{MR2172488, garoufalidis2005colored} of the corresponding knot $K$,
\begin{align}
    \hat A(K;\hat x,\hat y,q)F_K(x,q)=0~.
\end{align}
Based on \cite{willetts2020unification}, it is proposed in \cite{Gukov:2020lqm} that the limit $q \to \z_p = e^{2 \pi i / p}$ of $F_K$ leads to the $p$-th ADO polynomials \cite{Akutsu:1991xc}.

The development of $F_K$ is currently following that of quantum knot invariants although its definition is not given yet.
In \cite{Park:2019xey}, an extension of $\widehat Z$ and $F_K$ to arbitrary gauge group $G$ was studied.
Analogous to HOMFLY-PT polynomials and superpolynomials,  $(a,t)$-deformation
of $F_K$ are put forth in \cite{Ekholm:2020lqy}.

Following this path, in \S\ref{section:formulae}, we provide a closed-form expression of the $(a,t)$-deformed $F_K$ for double twist knots with positive braids in this paper.
Also, in \S\ref{ADO}, we derive closed-form formulae of $t$-deformed ADO polynomials.

\section{Knot homology and super-$A$-polynomial}

  The categorifications of quantum knot invariants shed new light on knot theory not only because knot homologies are more powerful than quantum invariants but also because they are functorial \cite{rasmussen2010khovanov}. Moreover, they are endowed with rich structural properties \cite{Wedrich:2016smm}, and various differentials give remarkable relationships among themselves \cite{Dunfield:2005si,Gukov:2011ry,Gorsky:2013jxa,rasmussen2016some}.  In particular, some relations between HOMFLY-PT and Kauffman homology have been uncovered in \cite{Gukov:2005qp, Nawata:2013mzx}. The rich structural properties also help us find closed-form expressions of Poincar\'e polynomials of cyclotomic type \cite{Habiro:2008} for various knots and links.

  In this section, we will show a new relationship between colored HOMFLY-PT and Kauffman homology for thin knots. Also, we investigate the effects of differentials on super-$A$-polynomials of $\SU$ and $\SO$-type.

\subsection{Isomorphism between colored HOMFLY-PT and Kauffman homology for thin knots } \label{section:change-of-variable}

It has been apparent that the rich structural properties become manifest if we introduce quadruple-gradings to HOMFLY-PT and Kauffman homology.
The quadruply-graded HOMFLY-PT homology $(\mathscr{H}^{\text{HOMFLY-PT}}_R(K))_{i,j,k,l}$ with $(a,q,t_r,t_c)$-gradings was introduced in \cite{Gorsky:2013jxa}.
With this grading, we can associate a $\delta$-grading to every generator $x$ of the $[r]$-colored HOMFLY-PT homology defined as,
$$
\delta(x):=a(x)+\frac{q(x)}{2}-\frac{t_r(x)+t_c(x)}{2}.
$$
A knot $K$ is called homologically-thin if all generators of $\mathscr{H}^{\text{HOMFLY-PT}}_{[r]}(K)$ have the same $\delta$-grading  equal to $\frac{r}{2}S(K)$ , where $S(K)$ is the Rasmussen $s$-invariant of $K$ \cite{rasmussen2010khovanov}. Otherwise, it is called homologically-thick. In general, thick knots possess more intricate structures \cite{Dunfield:2005si,Gukov:2011ry}.

In a similar manner, the structral properties of quadruply-graded Kauffman homology $(\mathscr{H}^{\text{Kauffman}}_R(K))_{i,j,k,l}$ were studied in \cite{Nawata:2013mzx}.
In the following discussion in this section, we also use the tilde versions, $\widetilde{\mathscr{H}}^{\text{HOMFLY-PT}}_{[r^s]}(K)$ \cite{Gorsky:2013jxa}, $\widetilde{\mathscr{H}}^{\text{Kauffman}}_{[r^s]}(K)$ \cite{Nawata:2013mzx}, and their corresponding Poincar\'{e} polynomials
\begin{align}
\widetilde{\mathscr{P}}_{[r^s]}(K;a,Q,t_r,t_c)&:=\sum_{i,j,k,l}a^i Q^j t_r^k t_c^l \text{dim} \left(\widetilde{\mathscr{H}}^{\text{HOMFLY-PT}}_{[r^s]}(K)\right)_{i,j,k,l}, \\
\widetilde{\mathscr{F}}_{[r^s]}(K;a,Q,t_r,t_c)&:=\sum_{i,j,k,l}a^i Q^j t_r^k t_c^l \text{dim} \left(\widetilde{\mathscr{H}}^{\text{Kauffman}}_{[r^s]}(K)\right)_{i,j,k,l}~,
\end{align}
where the gradings of $\scH$ and $\wt\scH$ are related by
\begin{align}
\mathscr{F}_{[r^s]}(K;a,q,t_r,t_c)&:=\widetilde{\mathscr{F}}_{[r^s]}(K;a,q^s,t_rq^{-1},t_cq)\label{triplykauff}~,\\
\mathscr{P}_{[r^s]}(K;a,q,t_r,t_c)&:=\widetilde{\mathscr{P}}_{[r^s]}(K;a,q^s,t_rq^{-1},t_cq)~.
\end{align}

It turns out that there are differentials, called universal and diagonal in \cite{Nawata:2013mzx}, that relate $[r]$-colored Kauffman and HOMFLY-PT homology. Furthermore,  the well-known isomorphism of the representations \cite{Nawata:2013mzx}
$$
\left(\mathfrak{so}_6,[r]\right)\simeq\left(\mathfrak{sl}_4,[r^2]\right),
$$
leads to an isomorphism between bi-graded homologies
\be\label{so6-sl4}
\scH_{\mathfrak{so}_6,[r]}(K)\cong \scH_{\mathfrak{sl}_4,[r^2]}(K)~.
\ee
For a thin knot $K_{\text{thin}}$, this can be stated in terms of the Poincar\'e polynomials \cite[(4.57)]{Nawata:2013mzx}
\be
\widetilde{\mathscr{F}}_{[r]}(K_{\text{thin}};a\!=\!q^5,Q\!=\!q,t_r\!=\!q^{-1},t_c\!=\!qt)=\widetilde{\mathscr{P}}_{[r^2]}(K_{\text{thin}};a\!=\!q^4,Q\!=\!q^2,t_r\!=\!q^{-1},t_c\!=\!qt)~.
\ee

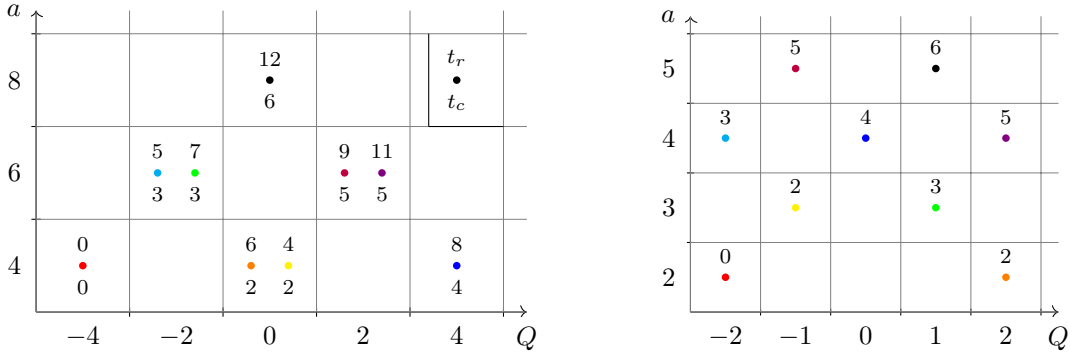
\begin{figure}[htbp]
\centering
\begin{tikzpicture}[scale=1.23]
\draw[->] (0,0) -- (5.25,0);
\draw[->] (0,0) -- (0,3.25);

\foreach \t/\ttext in {1/$-4$, 2/$-2$, 3/0, 4/2, 5/4} {
    \draw (\t, -0.05) -- (\t, 0.05);
    \node[below, outer sep=2pt, fill=white, font=\small]
        at (\t - 0.5, 0) {\ttext};
}
\node[below, outer sep=2pt, fill=white, font=\small] at (5.25,0) {$Q$};

\foreach \t/\ttext in {1/4, 2/6, 3/8} {
    \draw (-0.05, \t) -- (0.05, \t);
    \node[left, outer sep=2pt, fill=white, font=\small]
        at (0, \t - 0.5) {\ttext};
}
\node[left, outer sep=2pt, fill=white, font=\small] at (0,3.25) {$a$};

\draw[step=1cm, gray, very thin] (0,0) grid (5.25,3.25);

\node[inner sep=1pt, shape=circle, fill=red, label={[font=\scriptsize]above:$0$}, label={[font=\scriptsize]below:$0$}] at (0.5,0.5) {};
\node[inner sep=1pt, shape=circle, fill=blue, label={[font=\scriptsize]above:$8$}, label={[font=\scriptsize]below:$4$}] at (4.5,0.5) {};
\node[inner sep=1pt, shape=circle, fill=orange, label={[font=\scriptsize]above:$6$}, label={[font=\scriptsize]below:$2$}] at (2.3,0.5) {};
\node[inner sep=1pt, shape=circle, fill=yellow, label={[font=\scriptsize]above:$4$}, label={[font=\scriptsize]below:$2$}] at (2.7,0.5) {};
\node[inner sep=1pt, shape=circle, fill=cyan, label={[font=\scriptsize]above:$5$}, label={[font=\scriptsize]below:$3$}] at (1.3,1.5) {};
\node[inner sep=1pt, shape=circle, fill=green, label={[font=\scriptsize]above:$7$}, label={[font=\scriptsize]below:$3$}] at (1.7,1.5) {};
\node[inner sep=1pt, shape=circle, fill=purple, label={[font=\scriptsize]above:$9$}, label={[font=\scriptsize]below:$5$}] at (3.3,1.5) {};
\node[inner sep=1pt, shape=circle, fill=violet, label={[font=\scriptsize]above:$11$}, label={[font=\scriptsize]below:$5$}] at (3.7,1.5) {};
\node[inner sep=1pt, shape=circle, fill=black, label={[font=\scriptsize]above:$12$}, label={[font=\scriptsize]below:$6$}] at (2.5,2.5) {};

\node[inner sep=1pt, shape=circle, fill=black,label={[font=\scriptsize]above:$t_r$}, label={[font=\scriptsize]below:$t_c$}] at (4.5,2.5) {};
\draw (4.2, 2) -- (5, 2);
\draw (4.2, 2) -- (4.2, 3);

  \begin{scope}[shift={(7,0)},scale=0.75]
\draw[->] (0,0) -- (5.25,0);
\draw[->] (0,0) -- (0,4.25);

\foreach \t/\ttext in {1/$-2$, 2/$-1$, 3/0, 4/1, 5/2} {
    \draw (\t, -0.05) -- (\t, 0.05);
    \node[below, outer sep=2pt, fill=white, font=\small]
        at (\t - 0.5, 0) {\ttext};
}
\node[below, outer sep=2pt, fill=white, font=\small] at (5.25,0) {$Q$};

\foreach \t/\ttext in {1/2, 2/3, 3/4, 4/5} {
    \draw (-0.05, \t) -- (0.05, \t);
    \node[left, outer sep=2pt, fill=white, font=\small]
        at (0, \t - 0.5) {\ttext};
}
\node[left, outer sep=2pt, fill=white, font=\small] at (0,4.25) {$a$};

\draw[step=1cm, gray, very thin] (0,0) grid (5.25,4.25);

\node[inner sep=1pt, shape=circle, fill=red, label={[font=\scriptsize]above:$0$}] at (0.5,0.5) {};
\node[inner sep=1pt, shape=circle, fill=cyan, label={[font=\scriptsize]above:$3$}] at (0.5,2.5) {};
\node[inner sep=1pt, shape=circle, fill=orange, label={[font=\scriptsize]above:$2$}] at (4.5,0.5) {};
\node[inner sep=1pt, shape=circle, fill=yellow, label={[font=\scriptsize]above:$2$}] at (1.5,1.5) {};
\node[inner sep=1pt, shape=circle, fill=green, label={[font=\scriptsize]above:$3$}] at (3.5,1.5) {};
\node[inner sep=1pt, shape=circle, fill=blue, label={[font=\scriptsize]above:$4$}] at (2.5,2.5) {};
\node[inner sep=1pt, shape=circle, fill=violet, label={[font=\scriptsize]above:$5$}] at (4.5,2.5) {};
\node[inner sep=1pt, shape=circle, fill=purple, label={[font=\scriptsize]above:$5$}] at (1.5,3.5) {};
\node[inner sep=1pt, shape=circle, fill=black, label={[font=\scriptsize]above:$6$}] at (3.5,3.5) {};
\end{scope}
\end{tikzpicture}
\caption{An isomorphism between $[1^2]$-colored HOMFLY-PT homology (left) and uncolored Kauffman homology (right) for the trefoil. Generators with the same color are identified under the grading change \eqref{homology}. }\label{fig:isomorphism}
\end{figure}

Moreover, we find an isomorphism between $[1^2]$-colored HOMFLY-PT homology and the uncolored Kauffman homology of the trefoil and the figure-eight. For example, there is a one-to-one correspondence between their generators with the same color, shown in Figure \ref{fig:isomorphism}.
This motivates us to uplift \eqref{so6-sl4} to an isomorphism between HOMFLY-PT and Kauffman homologies in the case of a thin knot $K_{\text{thin}}$ with grading change
\begin{align}\label{homology}
\left(\widetilde{\mathscr{H}}^{\text{Kauffman}}_{[r]}(K_{\text{thin}})\right)_{i,j,k,l}\cong\left(\widetilde{\mathscr{H}}^{\text{HOMFLY-PT}}_{[r^2]}(K_{\text{thin}})\right)_{\frac32i+\frac34j-\frac12k,-\frac52i-\frac54j+\frac32k,i+\frac12j,l} ,
\end{align}
up to some overall grading shifts proportional to $\frac{rS(K)}{2}$ in $(a,Q,t_r)$ degrees.
This can be expressed in terms of the Poincar\'e polynomials
\be\label{kauffcov2}
\widetilde{\mathscr{F}}_{[r]}(K_{\text{thin}};a,Q,t_r,t_c)=\left(\frac{Q^3}{a t_r^2}\right)^{\frac{rS(K)}{2}}\widetilde{\mathscr{P}}_{[r^2]}(K_{\text{thin}};a^{\frac32}Q^{-\frac52}t_r,a^{\frac34}Q^{-\frac54}t_r^{\frac12},a^{-\frac12}Q^{\frac32},t_c)~.
\ee
We would like to emphasize that the isomorphism holds only for thin knots, but it is not true for thick knots. For thick knots, the dimensions of $[r^2]$-colored quadruply-graded HOMFLY-PT homology and $[r]$-colored quadruply-graded Kauffman homology are different \cite[Appendix B]{Gukov:2011ry}.

Remark that the relation $(\ref{kauffcov2})$ only provides the method to obtain the Poincar\'e polynomial of $[r]$-colored Kauffman homology from that of $[r^2]$-colored HOMFLY-PT homology for a thin knot, but it does not allow us to use this relation in the opposite direction since the grading shifts of variables in (\ref{kauffcov2}) are not linearly independent.

In \cite{Kononov:2016cwp, Kameyama:2019mdf}, the Poincar\'e polynomial of quadruply-graded HOMFLY-PT homology of the double twist knot $K_{m,n}$ colored by rectangular Young tableau $[r^s]$ has been given. (See Figure \ref{fig:double_twist_knot} for the double twist knot.)
Since the double twist knots are known to be homologically-thin, we can obtain the expression of the Poincar\'e polynomial of $[r]$-colored Kauffman homology for the double-twist knots thanks to the relation (\ref{kauffcov2}). To verify the relations (\ref{homology}) and (\ref{kauffcov2}), we have checked that the colored Kauffman homology obtained by this substitution satisfies all the differentials and properties \cite{Nawata:2013mzx} of the colored Kauffman homology for double twist knots.  We also attached the Mathematica files that show these properties on the arXiv page.

\subsection{Differential on super-$A$-polynomials of SO-type} \label{section:super-a-polynomial}
The study of the large color behaviors of colored Jones polynomials led us to the volume conjecture \cite{murakami2001colored, Gukov:2003na}. The generalization of the volume conjecture to the large color behavior of the Poincar\'e polynomial led to the so-called super-$A$-polynomial \cite{Fuji:2012nx,Nawata:2013mzx}. As briefly mentioned above, there are the differentials that relate $[r]$-colored Kauffman homology and HOMFLY-PT homology. In this subsection, we will investigate the effect of the differential on super-$A$-polynomials.

To this end, let us first recall the generalized volume conjecture \cite{Gukov:2003na,Gukov:2008jlx} and the super-$A$-polynomials \cite{Fuji:2012nx}. Here we use Poincar\'e polynomials of triply-graded homology $\mathscr{Q}_{[r]}(K;a,q,t)$ only with $t_r$-grading, also called the superpolynomial,  defined as
\be\label{superpolynomial}
\mathscr{Q}_{[r]}(K;a,q,t):=\mathscr{Q}_{[r]}(K;a,q,t_r=t,t_c=1)~,
\ee
where $\mathscr{Q}=\scP \textrm{ or }  \scF$ represents  Poincar\'e polynomials of either HOMFLY-PT or Kauffman homology.
If a knot $K$ satisfies the exponential growth property \cite{Gukov:2011ry,Wedrich:2016smm} (such as thin knots and torus knots)
$$\mathscr{Q}_{[r]}(K;a,q=1,t)=\Bigl[\mathscr{Q}_{[1]}(K;a,q=1,t)\Bigr]^r~,$$
it is conjectured \cite[Conjecture 4.4]{Kucharski:2017ogk} that the superpolynomial colored by a symmetric representation can be expressed as
\be \label{general-superpoly}
\mathscr{Q}_{[r]}(K;a,q,t)=\sum_{l_{1}+l_{2}+\cdots+l_{n}=r}\left[\begin{array}{c}
r \\
l_{1}, l_{2}, \ldots, l_{n}
\end{array}\right]_{q^2} a^{\sum_{i=1}^{n} a_{i} l_{i}} t^{\sum_{i=1}^{n} t_{i} l_{i}} q^{\sum_{i, j=1}^{n} Q_{i, j} l_{i} l_{j}+\sum_{i=1}^{n} q_{i} l_{i}} ~,
\ee
where $(a_i,q_{i} ,t_{i})$ are constants and $Q_{i, j}$ is a quadratic form. Here the $q$-multibinomial is defined by
$$
\left[\begin{array}{c}
r \\
l_{1}, l_{2}, \ldots, l_{n}
\end{array}\right]_{q}:=\frac{\left(q ; q\right)_{r}}{\left(q ; q\right)_{l_{1}}\left(q ; q\right)_{l_{2}} \cdots\left(q ; q\right)_{l_{n}}}~.
$$
In the large color limit
\bea\label{largecolorlimit}
q=e^{\hbar}\rightarrow 1,\quad r\rightarrow \infty,\quad a=\text{fixed},\quad  t=\text{fixed},
\quad x=q^{2r}=\text{fixed}~,
\eea
a superpolynomial asymptotes to the form
\be\label{leading2}
\mathscr{Q}_{[r]}(K;a,q,t)\overset{r\to\infty,\hbar\to0}{\underset{x=q^{2r}}{\sim}}\exp\left(\frac{1}{2\hbar}\int \log y\frac{dx}{x}+\cdots\right)
~,\ee
where the integral is carried out on the zero locus of classical super-A-polynomial
$$\label{zerolocus}
\mathscr{A}(K;x,y,a,t)=0.
$$
The ellipsis in (\ref{leading2}) represents the subleading terms that possess regular behaviors under the limit $\hbar\to 0$.

If  $\mathscr{Q}_{[r]}(K;a,q,t)$ is written as a summation only over $n$ variables $l_1,\cdots,l_n$ as in \eqref{general-superpoly}, then its behavior under the large color limit could be approximated as
\begin{equation}\label{saddle1}
\mathscr{Q}_{\left[r\right]}(K;a,q,t) \sim \int e^{\frac{1}{2\hbar}\left(\widetilde{\scW}(K;z_1,\cdots,z_n;x,a,t)+\mathcal{O}(\hbar)\right)}dx\prod_{i=1}^{n}dz_i ~,
\end{equation}
where $z_i=q^{2l_i}$, $x=q^{2r}$.
The leading asymptotic behavior (\ref{leading2}) with respect to $\hbar$ comes from the saddle points
\be\label{saddle2}
\exp\left(z_i\frac{\partial\widetilde{\scW}(K;z_1,\cdots,z_n;x,a,t)}{\partial z_i}\right)\bigg|_{z_i=z_i^*}=1
~.\ee
and the zero locus of the classical super-A-polynomial is determined by
\be
\exp\left({x\frac{\partial \widetilde{\scW}(K;z_1^*,\cdots,z_n^*;x,a,t)}{\partial x}}\right)=y
~.\ee

Now let us consider the relationship between super-$A$-polynomials of $\SU$-type and $\SO$-type that comes from the universal differential.
According to \cite[\S4.5]{Nawata:2013mzx}, the universal differential $d^{\text{univ}}_{\to}$ has $(a,q,t)$-degree (0,2,1), which relates the $[r]$-colored Kauffman homology and the $[r]$-colored HOMFLY-PT homology as
\begin{equation}\label{trikauhom}
\mathscr{F}_{[r]}(K;a,q,t)\!=\!q^{rS(K)}\mathscr{P}_{[r]}(K;a q^{-1},q,t)+(1+q^2t)f(a,q,t)~.
\end{equation}
This implies
\begin{equation}\label{relation}
    \mathscr{F}_{[r]}(K;a,q,-q^{-2})=q^{rS(K)}\mathscr{P}_{[r]}(K;a q^{-1},q,-q^{-2}) ~.
\end{equation}
For a knot satisfying the exponential growth property, the superpolynomial admits an expression \eqref{general-superpoly}. Therefore, it is straightforward to verify that the leading order of the large color asymptotic behavior of the right hand side of \eqref{relation} is the same as that of the colored HOMFLY-PT polynomial although the change of variables $a\to a q^{-1},t=-q^{-2}$ is non-trivial. As a result, we have the large color asymptotic behavior
\be
 \mathscr{F}_{[r]}(K;a,q,-q^{-2}) \sim \int e^{\frac{1}{2 \hbar}\left(\widetilde{\cW}^{\SU}\left(K ; z_{i} ; x, a\right)+\mathcal{O}(\hbar)\right)} d x d z_{i}~.
 \ee
Thus, it encodes the information about the $a$-deformed $A$-polynomial $A^{\textrm{SU}}(K;x,y,a)$ of SU-type. On the other hand, the decategorification limit $t=-1$ is just the colored Kauffman polynomial $\mathscr{F}_{[r]}(K;a,q,-1)$ and its large color limit yields $a$-deformed $A$-polynomial  $A^{\textrm{SO}}(K;x,y,a)$ of SO-type. Therefore, the super-$A$-polynomial $\mathscr{A}^{\textrm{SO}}(K;x,y,a,t )$ of SO-type contains both $A^{\textrm{SO}}(K;x,y,a)$ and $A^{\textrm{SU}}(K;x,y,a)$ at the $t=-1$ specialization.

\begin{conjecture}\label{A-SO-SU}
For a knot satisfying the exponential growth property, the zero locus of the super-$A$-polynomial of SO-type at $t=-1$
$$\mathscr{A}^{\textrm{SO}}(K;x,y,a,t=-1 )=0$$
includes two branches $A^{\textrm{SO}}(K;x,y,a)=0$ and $A^{\textrm{SU}}(K;x,y,a)=0$.
\end{conjecture}

Let us take the trefoil as an example in which the super-$A$-polynomial of $\SO$-type  can be calculated. The colored superpolynomial of the trefoil \cite[(5.8)]{Nawata:2013mzx} and the corresponding twisted superpotential \cite[(6.15)]{Nawata:2013mzx} can be written as
\bea\label{triple-Kauffman-trefoil}
&\scF_{[r]}({\bf 3_1}; a,q,t)\cr
&=\sum_{k=0}^{r}\sum_{j=0}^{k}\sum_{i=0}^{r-k} a^{i - k + 3 r}q^{3 k - 2 j (1 + r) + r (2 r-3) + i (2 j + 2 r-1)} t^{ 2 (i - j + r)}   \left[\begin{array}{c}r\\ k \end{array}\right]_{q^2}\left[\begin{array}{c}k \\ j \end{array}\right]_{q^2}\left[\begin{array}{c}r-k\\ i \end{array}\right]_{q^2}\cr
 &\hspace{3cm}\times(-  a^2 t^3 q^{ 2 r-2}  ;q^2)_j  (-q^2 t;q^2)_{r-k} (- a q^{-1} t;q^2)_i~.
\eea
and
\bea
\widetilde{\scW}^{\textrm{SO}}({\bf 3_1};x,w,v,z,a,t)&=\log\left(\tfrac{wx^3}{z}\right)\log a-\log v\log x+(\log x)^2+\log w\log (vx)\cr
		&+2\log\left(\tfrac{wx}{v}\right)\log t-\tfrac{3\pi^{2}}{6}-\Li_{2}(x)+\Li_{2}(v)+\Li_{2}(zv^{-1})\cr
		&+\Li_{2}(w)+\Li_{2}(xz^{-1}w^{-1})+\Li_{2}(- a^{2}t^{3}x)-\Li_{2}(- a^{2}t^{3}xv)\cr
		&+\Li_{2}(-t)-\Li_{2}(-txz^{-1})+\Li_{2}(- a t)-\Li_{2}(- a tw)~,
		\label{W1trefoil}
\eea
where we introduce the variables as $w=q^{2i}$, $v=q^{2j}$ and $z=q^{2k}$. Obviously, the $t=-1$ specialization leads to the large color limit of colored Kauffman polynomial \be\label{W-SO}\widetilde{\scW}^{\textrm{SO}}({\bf 3_1};x,w,v,z,a,t=-1)=\widetilde{\cW}^{\textrm{SO}}({\bf 3_1};x,w,v,z,a)~.\ee On the other hand, at the $t=-q^{-2}$ specialization, the expression \eqref{triple-Kauffman-trefoil} collapses to $k=r$ and $i=0$ due to the term $(-q^2 t;q^2)_{r-k}$ in the summand so that it is written by the summation only over $j$. Correspondingly, the twisted superpotential encodes the large color limit of colored HOMFLY-PT polynomial
\be\label{W-SU}\widetilde{\scW}^{SO}({\bf 3_1};x,w=1,v,z=x,a,t=-1)=\widetilde{\cW}^{\textrm{SU}}({\bf 3_1};x,v,a)~.\ee
The saddle point equations \eqref{saddle1} and \eqref{saddle2} yield the classical super-$A$-polynomial of $\SO$-type of the trefoil \cite[(6.19)]{Nawata:2013mzx}, and its $t=1$ specialization can be consequently written as
\begin{align}\label{soapoly311}
  &\mathscr{A}^{\SO}(\mathbf{3_1};x,y,a,t=-1)
  =(1-ax)A^{\SO}(\mathbf{3_1};x,y,a)A^{\SU}(\mathbf{3_1};x,y,a)~,
\end{align}
where
\begin{align}\label{suapoly312}
     A^{\SO}(\mathbf{3_1};x,y,a)&=a^7x^7-a^7x^6-a^2xy+y~,\\
   A^{\SU}(\mathbf{3_1};x,y,a)&=y^2-ya^2+xya^2-2x^2ya^2-xy^2a^2-x^3a^4+x^4a^4+2x^2ya^4+x^3ya^4-x^4ya^6.\nonumber
\end{align}
(For $A^{\SU}(\mathbf{3_1};x,y,a)$, see \cite[(2.25)]{Fuji:2012nx} with $a\to a^2$.) It is evident that it satisfies Conjecture \ref{A-SO-SU}.

In principle, we can compute super-$A$-polynomials of SO-type for the double twist knots using the result in \S\ref{section:change-of-variable}. However, it is beyond the computational capabilities of current desktops to solve the saddle point equations \eqref{saddle1} and \eqref{saddle2} for them. However, since the colored Kauffman homology of a double twist knot is endowed with the universal differential $d^{\text{univ}}_{\to}$, we can easily check the relations corresponding to \eqref{W-SO} and \eqref{W-SU} at the level of the twisted superpotentials for the double twist knots. In this way, we can confirm Conjecture \ref{A-SO-SU} for the double twist knots.

Note that the diagonal differentials \cite[\S 4.6]{Nawata:2013mzx} also relate colored Kauffman and HOMFLY-PT homology. However, they do not bring us any interesting results once we take large color limits.

\section{$\SO(N)$ quantum $6j$-symbols} \label{section:6j-symbols}

$6j$-symbols are ubiquitous in physics and mathematics \cite{landau1958quantum,1965Quantum,biedenharn1981angular,butler1981point,varshalovich1988quantum}, and they reveal very rich symmetries. In particular, quantum knot invariants can be evaluated by using quantum $6j$-symbols \cite{Kirillov:1991ec,Moore:1988qv,Moore:1989vd}. Despite their importance and versatility, it is difficult to evaluate them in higher ranks. Obtaining a closed-form expression is moreover indispensable to study their symmetries as in \cite{1965Quantum, Alekseev:2019nzw}.

 Kauffman polynomials (homology) colored by symmetric representations for non-torus knots obtained in \S\ref{section:change-of-variable} in principle encode the information about $\SO(N)$ quantum $6j$-symbols. Hence, we attempt to extract the information in this section. First, we quantize $\SO(N)$ $6j$-symbols with all symmetric representations obtained by  Ali\v{s}auskas \cite{alisauskas1987some, Alisauskas_2002}. Next, we give some evidence to its validity by using representation theory and computing Kauffman polynomials.

\subsection{$\SO(N)$ quantum $6j$-symbols for symmetric representations}

Multiplicity-free $6j$-symbols of a Lie group $G$ take the form
$$
\begin{Bmatrix}
R_1 & R_2 & R_{12} \\
R_3 & R_4 & R_{23}
\end{Bmatrix}_{G}~,
$$
where the arguments are representations of $G$, and they obey the fusion rule, $R_{12} \in (R_1 \otimes R_2) \cap  (\overline R_3 \otimes \overline  R_4) $,
$R_{23} \in (R_2 \otimes R_3) \cap  (\overline R_1 \otimes \overline  R_4)$.

Ali\v{s}auskas has obtained the closed-form expression for classical $\SO(N)$ $6j$-symbols with all representations symmetric \cite{alisauskas1987some, Alisauskas_2002}.  Kirillov and Reshetikhin naturally quantized the classical $6j$-symbols of $\SU(2)$ \cite{Kirillov:1991ec} by bringing the hypergeometric series ${}_4F_3$ to ${}_4\varphi_3$. Therefore, we conjecture that we can naturally quantize Ali\v{s}auskas' results
to a closed-form expression of quantum $\SO(N)$ ($N \geq 4$) $6j$-symbols for all symmetric representations,
\begin{equation}
\label{quantum6j}
    \begin{aligned}
        \begin{Bmatrix}
        a & b & e \\
        d & c & f
        \end{Bmatrix}_{\SO(N)}=&\left(\frac{[2 c+N-2]_q[2 d+N-2]_q[2 e+N-2]_q}{[2]_q^3 \dim_q[c]\dim_q[d]\dim_q[e]}\right)^{1 / 2} \\
        & \times\begin{pmatrix}
        c & d & e \\
        0 & 0 & 0
        \end{pmatrix}_{N}^{-1} \sum_{l^{\prime}}(-1)^{(c+d-e) / 2+f+N+l^{\prime}}\left[2 l^{\prime}+N-3\right]_q \\
        & \times\begin{Bmatrix}
       \frac{1}{2} b & \frac{1}{2} f+\frac{1}{4} N-1 &\frac{1}{2} d+\frac{1}{4} N-1 \\
        \frac{1}{2} f+\frac{1}{4} N-1 & \frac{1}{2}(b+N)-2 & l^{\prime}+\frac{1}{2} N-2
        \end{Bmatrix}_{\SU(2)} \\
        & \times\begin{Bmatrix}
        \frac{1}{2} a & \frac{1}{2} f+\frac{1}{4} N-1 & \frac{1}{2} c+\frac{1}{4} N-1 \\
          \frac{1}{2} f+\frac{1}{4} N-1 &   \frac{1}{2}(a+N)-2 &   l^{\prime}+\frac{1}{2} N-2
        \end{Bmatrix}_{\SU(2)} \\
        & \times\begin{Bmatrix}
       \frac{1}{2} a & \frac{1}{2} b+\frac{1}{4} N-1 &  \frac{1}{2} e+\frac{1}{4} N-1 \\
        \frac{1}{2} b+\frac{1}{4} N-1 & \frac{1}{2}(a+N)-2 & l^{\prime}+\frac{1}{2} N-2
        \end{Bmatrix}_{\SU(2)}\\
        & \times\left(\frac{[l^{\prime}]_q![N-3]_q!}{\left[l^{\prime}+N-4\right]_q!}\right)^{1 / 2},
        \end{aligned}
\end{equation}
where the curly braces on the right hand side are the quantum $6j$-symbols of $\SU(2)$ \cite{Kirillov:1991ec}.
Since we are only dealing with all symmetric representations, $a, b, c, d, e, f$ are the number of boxes in the corresponding Young tableau of one row. Here
a quantum number is defined as
\be \label{def:qno}
[x]_q = \frac{q^x-q^{-x}}{q-q^{-1}}~.
\ee
Also, a $q$-factorial is defined as the product of quantum numbers
\be
[x]_q! = [x]_q [x-1]_q \cdots [x - \lfloor x \rfloor]_q~,
\ee
where $\lfloor x \rfloor$ is the floor of $x$ so that the definition
of $q$-factorial works for
both integer and half-integer arguments.
The quantum dimension of the $\SO(N)$ symmetric representation $[l]$ is given by
\be
\dim_q[l] = \frac{[l+N-3]_q!}{[l]_q! [N-2]_q!}
\left([l+N-2]_q+[l]_q\right),
\ee
and the special quantum $3j$-symbols are defined as
\begin{equation}
    \begin{aligned}
        \begin{pmatrix}
        l_{1} & l_{2} & l_{3} \\
        0 & 0 & 0
        \end{pmatrix}_{N} =& \frac{1}{[N / 2-1]_q!}\left(\frac{[J+N-3]_q !}{[N-3]_q ! [J+N / 2 - 1]_q!}\right.\\
        &\left.\times \prod_{i=1}^{3} \frac{\left[2l_{i}+N -2\right]_q \left[J-l_{i}+N / 2-2\right]_q!}{[2]_q \dim_q[l_i]\left[J-l_{i}\right]_q !}\right)^{1 / 2},
        \end{aligned}
\end{equation}
where $J=\frac{1}{2}(l_1+l_2+l_3)$.

Note that $R_{12}$, $R_{23}$ can be non-symmetric representations even when $R_1 = R_2 = R_3 = R_4$ are symmetric representations. (Some fusion rules can be found in \eqref{SON-fusion-rule}.)
Furthermore, there are several relations from representation theory, which would lead to identities among the quantum $6j$-symbols.
\begin{itemize}
    \item The isomorphism of representations, $(\mathfrak{so}_6, [r]) \simeq (\mathfrak{sl}_4,[r,r])$, would lead to equations,
    \begin{equation}
        \begin{Bmatrix}
        [r] & [r] & [2l-j,j] \cr
        [r] & [r] & [2l^\prime - j^\prime, j^\prime]
        \end{Bmatrix}_{\SO(6)} =  \begin{Bmatrix}
        [r,r] & [r, r] & [2l,2l-j,j] \cr
        [r,r] & [r, r] & [2l^\prime,2l^\prime - j^\prime, j^\prime]
        \end{Bmatrix}_{\SU(4)},
    \end{equation}
    where $0 \le l,l^\prime \le r$, $0 \le j \le l$, $0 \le j^\prime \le l^\prime$.
    We have checked that this property holds using $\SU(4)$ quantum $6j$-symbols in \cite{Mironov:2016coi}, and the result in the next subsection.
    Since now we only have the closed form expression of $\SO(N)$ quantum $6j$-symbols for symmetric representations,
    by setting $j = j^\prime = 0$, the above equation reduces to
    \begin{equation}
         \begin{Bmatrix}
        [r] & [r] & [2l] \cr
        [r] & [r] & [2l^\prime]
        \end{Bmatrix}_{\SO(6)} =  \begin{Bmatrix}
        [r,r] & [r,r] & [2l,2l] \cr
        [r,r] & [r,r] & [2l^\prime,2l^\prime]
        \end{Bmatrix}_{\SU(4)},
    \end{equation}
    where $0\le l \le r$. Thus we can easily obtain $\SU(4)$ quantum $6j$-symbols of such type, using our closed formula (\ref{quantum6j}).

    \item Also, there is an isomorphism between Lie algebras, $\mathfrak{so}_4 \simeq \mathfrak{sl}_2 \oplus \mathfrak{sl}_2$. Restricting ourselves to the case where all representations are symmetric, we have
    \begin{equation}
        \begin{Bmatrix}
        [r] & [r] & [2l] \cr
        [r] & [r] & [2l^\prime]
        \end{Bmatrix}_{\SO(4)}
        =
       \left( \begin{Bmatrix}
        [r] & [r] & [2l] \cr
        [r] & [r] & [2l^\prime]
        \end{Bmatrix}_{\SU(2)}\right)^2,
    \end{equation}
    where $0 \le l, l^\prime \le r$.

    \item Although our formula only works for $\SO(N)$ ($N \ge 4$), it's still worthy to note that there is another isomorphism of representations, $(\mathfrak{so}_3, [r]) \simeq (\mathfrak{sl}_2, [2r])$. This leads to
    \begin{equation}
        \begin{Bmatrix}
        [r] & [r] & [2l] \cr
        [r] & [r] & [2l^\prime]
      \end{Bmatrix}_{\SO(3)}\bigg|_{q\to q^2}
        =
        \begin{Bmatrix}
        [2r] & [2r] & [4l] \cr
        [2r] & [2r] & [4l^\prime]
        \end{Bmatrix}_{\SU(2)},
    \end{equation}
    where $0 \le l, l^\prime \le r$.
\end{itemize}

\subsection{$\SO(N)$ fusion matrices for symmetric representations}
Since the fusion rule of $\SO(N)$ symmetric representations is given by
\be\label{SON-fusion-rule}
[r] \otimes [r] =\bigoplus_{l=0}^{r} \bigoplus_{j=0}^{l}[2 l-j, j]~,
\ee
we need full $\SO(N)$ fusion matrices including the case that $R_{12}$ and $R_{23}$ are non-symmetric to evaluate Kauffman polynomials colored by symmetric representations,.
At a root of unity $q$, the WZNW fusion matrix and the corresponding quantum $6j$-symbol are related by
\be
\label{6j}
a_{R_{12} R_{23}}\left[\begin{array}{cc}
    R_{1} & R_{2} \\
    R_{3} & R_{4}
    \end{array}\right]=\epsilon_{\left\{R_{i}\right\}} \sqrt{\operatorname{dim}_{q} {R_{12}} \operatorname{dim}_{q} {R_{23}}}\begin{Bmatrix}
    R_{1} & R_{2} & R_{12} \\
    R_{3} & R_{4} & R_{23}
    \end{Bmatrix},
\ee
where $\epsilon_{\{R_i\}} = \pm 1$ and $\operatorname{dim}_{q} R$ is the quantum dimension of the representation $R$.
Since $\SO(N)$ representations are real, for colored knot invariants,
we only need to consider fusion matrix of the type
$a_{ts}\begin{bmatrix}
    \begin{smallmatrix}
        R & R\\ R & R
    \end{smallmatrix}
\end{bmatrix}$,
where $t,s \in R \otimes R$.

To obtain full fusion matrices,
we can make use of their properties and symmetries \cite{1965Quantum}
\begin{itemize}
    \item The fusion matrix is symmetric,
    \be
    a_{ts} \begin{bmatrix}R & R \\ R & R\end{bmatrix} =
    a_{st} \begin{bmatrix}R & R \\ R & R\end{bmatrix}.
    \ee

    \item The fusion matrix is orthonormal,
    \be
    \sum_s
    a_{ts} \begin{bmatrix}R & R \\ R & R\end{bmatrix}
    a_{sm} \begin{bmatrix}R & R \\ R & R\end{bmatrix} =
    \delta_{tm}.
    \ee

    \item The entries in the row and column, which corresponds to the trivial representation, are known,
    \be\label{trivialrep}
    a_{\emptyset s} \begin{bmatrix}R & R \\ R & R\end{bmatrix} = \e_s^{R}\frac{\sqrt{\operatorname{dim}_q s}}{\operatorname{dim}_q R},
    \ee
    where $\emptyset$ is the trivial representation.
\end{itemize}
Note that $6j$-symbols enjoy other symmetries, such as the Racah identity and the pentagon (Biedenharn-Elliot) identity, which we do not use in this paper.
We already know the entries when $t$ and $s$ are symmetric representations (\ref{quantum6j}), and the entries corresponding to the trivial representation (\ref{trivialrep}).
Then we can solve the rest entries by using the fact that the fusion matrix is symmetric and orthonormal.
In this paper, we will present our results of the $\SO(N)$ fusion matrices
$a_{ts}\begin{bmatrix}
    \begin{smallmatrix}
        R & R\\ R & R
    \end{smallmatrix}
\end{bmatrix}$
for $R= ~$\yng(1)~, \yng(2).

\bigskip
1. $R$ = \yng(1)
    \begin{displaymath}
        $${$\begin{Bmatrix}
        R & R & t\\
        R & R & s
        \end{Bmatrix}=\frac{1}{\operatorname{dim}_q R}$
        $\left({
        \begin{tabular}{c|c c c}
        {}
        & $s=\emptyset \quad$
        & ${\Yvcentermath1\Yboxdim6pt\yng(1,1)}$
        & ${\Yvcentermath1\Yboxdim6pt\yng(2)}$ \\
        \hline
        {$t=\emptyset$}
        & $1$
        & $1$
        & $1$
        \\
         {${\Yvcentermath1\Yboxdim6pt\yng(1,1)}$}
        & $1$
        & $\frac{[N]_q}{[N-1]_q([N-2]_q+[2]_q)}$
        & $-\frac{1}{[N-1]_q}$
        \\
        {${\Yvcentermath1\Yboxdim6pt\yng(2)}$}
        & $1$
        & $-\frac{1}{[N-1]_q}$
        & $\frac{[N-2]_q}{[N-1]_q([N]_q+[2]_q)}$\\
          \end{tabular} }\right)$}~,
        $$
    \end{displaymath}
    where the quantum dimension $\operatorname{dim}_q R = {[N-1]_q+1}$.

\bigskip
2. $R$ = \yng(2)
    \begin{displaymath}
        $${$\begin{Bmatrix}
        R & R & t\\
        R & R & s
        \end{Bmatrix}=\frac{1}{\operatorname{dim}_q R}$
        $\left({\renewcommand\tabcolsep{5.0pt}
        \begin{tabular}{c|c c c c c c}
        {}
        & $s=\emptyset$
        & ${\Yvcentermath1\Yboxdim6pt\yng(1,1)}$
        & ${\Yvcentermath1\Yboxdim6pt\yng(2)}$
        & ${\Yvcentermath1\Yboxdim6pt\yng(2,2)}$
        & ${\Yvcentermath1\Yboxdim6pt\yng(3,1)}$
        & ${\Yvcentermath1\Yboxdim6pt\yng(4)}$\\
        \hline
        {$t=\emptyset$} & $1$ & $1$ & $1$ & $1$ & $1$ & $1$\\
        {${\Yvcentermath1\Yboxdim6pt\yng(1,1)}$}
        & $1$ & $b_{11}$ & $b_{12}$ & $b_{13}$ & $b_{14}$ & $b_{15}$\\
        {${\Yvcentermath1\Yboxdim6pt\yng(2)}$}
        & $1$ & $b_{21}$ & $b_{22}$ & $b_{23}$ & $b_{24}$ & $b_{25}$\\
        {${\Yvcentermath1\Yboxdim6pt\yng(2,2)}$}
        & $1$ & $b_{31}$ & $b_{32}$ & $b_{33}$ & $b_{34}$ & $b_{35}$\\
        {${\Yvcentermath1\Yboxdim6pt\yng(3,1)}$}
        & $1$ & $b_{41}$ & $b_{42}$ & $b_{43}$ & $b_{44}$ & $b_{45}$\\
        {${\Yvcentermath1\Yboxdim6pt\yng(4)}$}
        & $1$ & $b_{51}$ & $b_{52}$ & $b_{53}$ & $b_{54}$ & $b_{55}$\\
          \end{tabular} }\right)$}~,
        $$
    \end{displaymath}
    where the quantum dimension $\operatorname{dim}_qR = [N-1]_q + [N-1]_q [N]_q/[2]_q$, and
    \begin{small}
    \begin{align}\nonumber
        & b_{11} = \frac{-q^4-q^N-q^{2+N}+q^{4+N}-q^{2+2N}+q^{4+2N}+q^{6+2N}+q^{2+3N}}{(1+q^2)(-1+q^N)(q^4+q^N+q^{2N}+q^{4+N})}, \cr
        & b_{12} = b_{21} = \frac{q^{N+2}+1}{q^{N+2}+q^N+q^2+1}, \cr
        & b_{13} = b_{31} = \frac{ q^N \left(q^2-1\right)\left(q^{N+2}+1\right)}{q^{3 N}+q^{2 N+4}-q^N-q^4}, \cr
        & b_{14} = b_{41} = \frac{\left(q^2-1\right)^2 q^N}{q^{2 N}+q^{N+4}+q^N+q^4}, \cr
        & b_{15} = b_{51} = \frac{\left(q^4-1\right) q^{N-2}}{1-q^{2 N}}, \cr
        & b_{22} = \frac{-q^{2 N}\!+\!q^{N+2}\!+\!3 q^{N+4}\!-\!2 q^{N+6}\!-\!q^{N+8}\!-\!2 q^{2 N+2}\!+\!3 q^{2 N+4}\!+\!q^{2 N+6}\!-\!2 q^{2 N+8}\!+\!q^{3
        N+6}\!-\!2 q^N\!+\!q^2}{\left(q^2+1\right) \left(q^N+1\right) \left(q^N-q^2\right) \left(q^{N+4}-1\right)}, \cr
        & b_{23} = b_{32} = -\frac{\left(q^2-1\right) q^N}{\left(q^N+1\right) \left(q^N-q^2\right)}, \cr
        & b_{24} = b_{42} = \frac{q^N \left(q^2-1\right)^2  \left(q^{N+2}+1\right)}{\left(q^2-q^N\right) \left(q^{N+4}+q^{2
        N+4}-q^N-1\right)}, \cr
        & b_{25} = b_{52} = \frac{\left(q^4-1\right) q^N}{q^{N+4}+q^{2 N+4}-q^N-1}, \cr
        & b_{33} = \frac{ q^{2 N}\left(q^2\!-\!1\right)^2 \left(q^2\!+\!1\right) \left(q^{2 N}\!-\!q^{N\!+\!2}\!+\!q^{N\!+\!6}\!-\!q^{2 N\!+\!2}\!-\!2 q^{2
        N\!+\!4}\!-\!q^{2 N\!+\!6}\!+\!q^{2 N\!+\!8}\!+\!q^{3 N\!+\!2}\!-\!q^{3 N\!+\!6}\!+\!q^{4 N\!+\!4}\!+\!q^4\right)}{\left(q^N\!-\!1\right)
        \left(q^N\!+\!1\right) \left(q^N\!-\!q^2\right) \left(q^N\!+\!q^4\right) \left(q^{2 N}\!-\!q^6\right) \left(q^{2
        N+2}\!-\!1\right)}, \cr
        & b_{34} = b_{43} = -\frac{q^{2 N}\left(q^2-1\right)^2 \left(q^2+1\right)  \left(q^{N+2}+1\right)}{\left(q^N+1\right)
        \left(q^{2 N}-q^{N+2}+q^{N+4}-q^6\right) \left(q^{2 N+2}-1\right)}, \cr
        & b_{35} = b_{53} = \frac{\left(q^2-1\right)^2 \left(q^2+1\right) q^{2 N-2}}{\left(q^{2 N}-1\right) \left(q^{2
        N+2}-1\right)}, \cr
        & b_{44} = \frac{q^{2 N}\left(q^2-1\right)^2 \left(q^2+1\right)  \left(2 q^{N+2}-2 q^{N+4}+q^{N+6}+q^{2
        N+4}-q^N-q^2\right)}{\left(q^{2 N}-q^{N+2}+q^{N+4}-q^6\right) \left(q^{2 N+2}-1\right)
        \left(q^{N+4}+q^{2 N+4}-q^N-1\right)}, \cr
        & b_{45} = b_{54} = -\frac{\left(q^2-1\right)^2 \left(q^2+1\right) q^{2 N}}{\left(q^{2 N+2}-1\right) \left(q^{N+4}+q^{2
        N+4}-q^N-1\right)}, \cr
        & b_{55} = \frac{q^{2 N+2}\left(q^2-1\right)^2 \left(q^2+1\right) \left(q^N-q^2\right)}{\left(q^{N+6}-1\right)
        \left(q^{2 N+2}-1\right) \left(q^{N+4}+q^{2 N+4}-q^N-1\right)}. \nonumber
    \end{align}
    \end{small}

Using these two $\SO(N)$ fusion matrices and braiding operator eigenvalues, we have computed colored Kauffman polynomials for double twist knots  when  $R\!=$\yng(1) , \yng(2). The results coincide with  $\mathscr{F}_{[r]}(K;a=q^{N-1},q,t=-1)$  obtained in \S\ref{section:change-of-variable}, which comes from the Poincar\'e polynomial of $[r^2]$-colored HOMFLY-PT homology \cite[(3.4)]{Kameyama:2019mdf} by using the relation (\ref{kauffcov2}).

\section{Deformed $F_K$ and ADO polynomials}\label{section:FK}

In the study of the 3d/3d correspondence, a new invariant $\widehat Z$ of a closed 3-manifold was proposed as a BPS $q$-series \cite{Gukov:2016gkn, Gukov:2017kmk}.
To incorporate TQFT methods and surgery formula, an analogous series $F_K$ for a knot complement $S^3\backslash K$ \cite{Gukov:2019mnk} was introduced.
Although a mathematical definition of $\hat{Z}$ or $F_{K}$ has not been fleshed out yet, there are many approaches to understand them.
A recipe to compute $\wh Z$ for a plumbed 3-manifold with a negative definite linking matrix is given in \cite{Gukov:2017kmk}. We can also use resurgence to calculate $\wh Z$ or $F_K$  \cite{Gukov:2016njj,Gukov:2019mnk}. Also, for a positive braid knot $K$, $F_{K}(x, q)$ can be computed using the $R$-matrix for Verma modules \cite{Park:2020edg}. In particular,
closed-form expressions of $F_K$ for the double twist knots with positive braids are given in \cite{Park:2020edg}, based on the formulae of colored Jones polynomials \cite{lovejoy2017colored,lovejoy2019colored}. The expressions are compatible with Habiro's cyclotomic expansion \cite{Habiro:2008}.
In this section, we will generalize the results of  \cite{Park:2020edg} to the $(a,t)$-deformed case \cite{Ekholm:2020lqy}, and we will see by explicit computations that the ADO invariants \cite{Akutsu:1991xc,willetts2020unification,Gukov:2020lqm} are compatible with the cyclotomic expansions.

   \subsection{Deformed $F_K$ of double twist knots}\label{section:formulae}

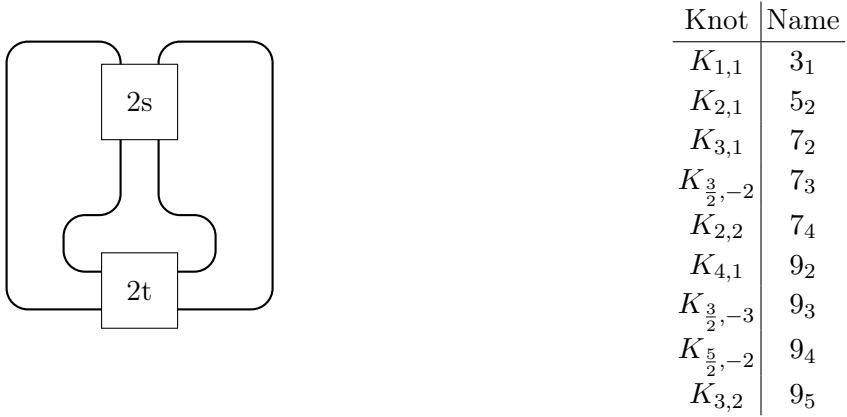
\begin{figure}[ht]
     \begin{minipage}[b]{8cm}\centering
   	\centering
   	\raisebox{-1.5cm}{\begin{tikzpicture}
   \draw[thick, rounded corners = 8pt]
       (-0.25, 0) -- (-0.25, -1) -- (-1, -1) -- (-1, -1.75) -- (-0.5, -1.75);
       \draw[thick, rounded corners = 8pt]
           (0.25, 0) -- (0.25, -1) -- (1, -1) -- (1, -1.75) -- (0.5, -1.75);
       \draw (-0.5, 0) rectangle (0.5,1);
       \draw (-0.5, -1.5) rectangle (0.5,-2.5);
       \draw[thick, rounded corners = 8pt]
       (0.25, 1.00) -- (0.25, 1.3) -- (1.75, 1.3) -- (1.75, -2.25) -- (0.5, -2.25);
       \draw[thick, rounded corners = 8pt]
       (-0.25, 1.00) -- (-0.25, 1.3) -- (-1.75, 1.3) -- (-1.75, -2.25) -- (-0.5, -2.25);
       \node (a) at (0,0.5){2s};
       \node (b) at (0,-2){2t};
   \end{tikzpicture}}
   	\label{fig:two_braid_knot}
   	\end{minipage}
   	  \begin{minipage}[b]{8cm}\centering
   	 $$ \begin{array}{c | c}\textrm{Knot}& \textrm{Name}\\ \hline{K_{1,1}}&{3 _{1}} \\{K_{2,1}}&{5 _{2}} \\{K_{3,1}}&{7 _{2}} \\{K_{\frac{3}{2},-2}}&{7 _{3}} \\{K_{2,2}}&{7 _{4}} \\{K_{4,1}}&{9 _{2}} \\{K_{\frac{3}{2},-3}}&{9 _{3}} \\{K_{\frac{5}{2},-2}}&{9 _{4}}\\{K_{3,2}}&{9 _{5}}
   	 \end{array}$$
   	  \end{minipage}
   	 \caption{Double twist knot $K_{s,t}$ and corresponding knots in Rolfsen table.}\label{fig:double_twist_knot}
   \end{figure}

   We study the double twist knots, $K_{s,t}~(s,t\in \frac{1}{2}\mathbb{Z})$, whose notation and the corresponding knots in Rolfsen table are shown in figure \ref{fig:double_twist_knot}. Among them, two classes possess only positive braids.

   \subsection*{Double twist knot $K_{m,n}~(m,n\in \mathbb{Z}_+)$}

   The cyclotomic expansions of the $[r]$-colored superpolynomials for the double twist knots $K_{m,n}~(m,n\in \mathbb{Z}_+)$ are given in \cite{Nawata:2012pg, Gukov:2015gmm}. In fact, by substitution $q^{2r}=x$ and changing the upper limit from $r$ to $\infty$, we can obtain
the corresponding $(a,t)$-deformed $F_K$:
   \begin{equation}\label{Kmn}
   F_{K_{m,n}}(x,a,q,t)=(-tx)^{\log_q(-at)-1}\sum_{k=0}^\infty
   g_{K_{m,n}}^{(k)} (x, a, q,t),
   \end{equation}
   where
   \begin{align}\label{eq:g}
   g_{K_{m,n}}^{(k)} (x, a, q, t) =~&
   \sum_{j=0}^k
   \frac{\left(x;q^{-1}\right)_k \left(-at/q;q\right)_k \left(-a x t^3;q\right)_j}
   {(q;q)_k}
   \cr
   &\times (-1)^j\, x^{k-j}\, a^{-j}\, q^k\, t^{-3j+2k}\begin{bmatrix}k\\j\end{bmatrix}_q
   {\rm Tw}_{K_{m,n}}^{(j)}(a,q,t).
   \end{align}
   The twist factor is defined as
   \begin{equation}\label{twist2}
   {\rm Tw}_{K_{m,n}}^{(j)}(a,q,t)=\sum_{l=0}^j (-1)^l\, q^{\frac{1}{2} l(l+1)-jl}\, {\rm tw}_m^{(l)}(a,q,t)\, {\rm tw}_n^{(l)}(a,q,t)\begin{bmatrix}j\\l\end{bmatrix}_q,
   \end{equation}
   with
   \begin{equation} \label{eq:tw}
     {\rm tw}_m^{(l)}(a,q,t)=\sum_{0\leq b_1 \leq \cdots \leq b_{m-1} \leq b_m =l}
   \prod_{i=1}^{m-1}\left(a^{b_i} q^{b_i (b_i-1)} t^{2 b_i}\right)
   {b_{i+1} \brack b_i}_q.
   \end{equation}
Note that, to connect the standard notation of $F_K$ and the ADO invariants, we rescale the variables $a\to a^{1/2}$ and $q\to q^{1/2}$ from the superpolynomial \eqref{superpolynomial}
$$
   F_{K_{m,n}}(x=q^r,a,q,t)=\scP_{[r]}(K_{m,n};a^{1/2},q^{1/2},t)~.
$$

   When we write the $(a, t)$-deformed $F_K$ as a four-variable series explicitly, we would usually omit the prefactor $(-t x)^{\log_q(-at) - 1}$.
   However, we have to keep it to verify the properties (\ref{tAlexander}), (\ref{property2}), (\ref{property3}) of $(a,t)$-deformed $F_K$.
   If we take $a = -t^{-1} q^N$, the prefactor simply reduces to $(-tx)^{N - 1}$.

   \subsection*{Double twist knot $K_{m+\frac{1}{2},-n}~(m,n \in \mathbb{Z}_+)$}
   Similarly, we can find the cyclotomic expansions of the $[r]$-colored superpolynomial for double twist knot $K_{m+\frac12, -n}~(m,n \in \mathbb{Z}_+)$ by using the structural properties of HOMFLY-PT homology. Then, we obtain the $(a, t)$-deformed $F_K$ for it:
   \begin{equation}\label{km+1/2}
   F_{K_{m+\frac{1}{2},-n}}(x,a,q,t)=(-tx^n)^{\log_q(-at) - 1}\sum_{k=0}^\infty
   g_{K_{m+\frac12,-n}}^{(k)}(x, a, q, t),
   \end{equation}
   where
   \begin{align}
   g^{(k)}_{K_{m+\frac12, -n}} (x, a, q, t) =~&
   \sum_{j=0}^k
   \frac{\left(x;q^{-1}\right)_k \left(-at/q;q\right)_k \left(-a x t^3;q\right)_j}
   {(q;q)_k}
   \cr
   &\times x^{k-j}\, \begin{bmatrix}k\\j\end{bmatrix}_q \, q^{\frac{1}{2}j(j-1)+k}\, t^{-j+2k}\,
   {\rm tw}_m^{(j)}(a,q,t) \, \tw_n^{(k)}(x,q,t).
   \label{eq:g2}
   \end{align}
   A new twist factor is defined as
   \begin{align} \label{eq:TW}
   \tw_n^{(k)}(x,q,t)
   =&\sum_{0\leq b_1 \leq \cdots \leq b_{n-1} \leq b_n =k}
   \prod_{i=1}^{n-1}\left(x^{2b_{i}}\, q^{-b_i (b_{i+1}-1)}\, t^{2 b_i}\right)
   {b_{i+1} \brack b_{i}}_q.
   \end{align}
   The prefactor becomes simply $(-tx^n)^{N - 1}$ when taking $a = -t^{-1} q^N$.

   \subsection*{Properties of $(a,t)$-deformed $F_K$}
   The formulae above for $F_K$ of the double twist knots become the usual $F_K$  in \cite{Park:2020edg} up on the specialization $a=q^2,t=-1$.
   If we further take $a = q^N, t=-1$, it reduces to $F_K^{\SU(N), sym}(x,q)$ \cite{Park:2019xey}, which is the $F_K$ with gauge group $G=\SU(N)$ associated with symmetric representations.

It is conjectured in \cite{Ekholm:2020lqy} that the $(a,t)$-deformed $F_K$ has the following property,
   \begin{equation}
       \hat{A}(K;\hat{x},\hat{y},a,q,t)F_K(x,a,q,t)=0.
   \end{equation}
   We have studied the classical super-$A$-polynomial in \S\ref{section:super-a-polynomial}, and $\hat{A}(K;\hat{x},\hat{y},a,q,t)$ is its quantum version \cite{Nawata:2012pg},
   which annihilates the colored superpolynomial of $K$.
   The operators in the quantum super-$A$-polynomial act on $(a,t)$-deformed $F_K$ as
   \begin{align}
       \hat x F_K(x,a,q,t)=x F_K(x,a,q,t),\qquad \hat y F_K(x,a,q,t)=F_K(x q,a,q,t).
   \end{align}
   In the case of $3_1,5_2,7_2$, the conjectural $(a,t)$-deformed $F_K$ can be annihilated by the corresponding super-$A$-polynomial in \cite{Nawata:2012pg}, the detail can be seen in the Mathematica file.

  It is proposed in \cite{Ekholm:2020lqy} that $(a,t)$-deformed $F_K$ satisfy the following properties:
   \begin{align}
       F_K (x, -t^{-1}, q, t) &= \Delta_K(x,t), \label{tAlexander}\\
       F_K (x, -t^{-1} q, q, t) &= 1, \label{property2}\\
      \lim_{q \to 1} F_K (x, -t^{-1} q^N, q, t) &= \frac{1}{\Delta_K(x,t)^{N - 1}}, \label{property3}
   \end{align}
   where $\Delta_K(x,t)$ is the $t$-deformed Alexander polynomial introduced in \cite{Ekholm:2020lqy},
   which can be obtained from the superpolynomial,
   \begin{equation}\label{alex-super}
       \Delta_K(x,t) = \mathscr{P}_{\yng(1)} (K;a^2 = - t^{-1}, ~ q^2 = x, ~ t).
   \end{equation}
We have checked that our closed formulae of the $(a,t)$-deformed $F_{K}$ for the double twist knots satisfy \eqref{tAlexander}.
Also, we give an analytical proof of \eqref{property2} and \eqref{property3} in Appendix \ref{Appendix1}.

\subsection{$t$-deformed ADO polynomials}\label{ADO}

As studied in \cite{Gukov:2020lqm}, at the limit $q\rightarrow \z_p = e^{\frac{2\pi i}{p}}$, $F_K(x,q)$ is equal to the $p$-th ADO polynomial \cite{Akutsu:1991xc} up to the Alexander polynomial $\Delta_K(x^p)$.
For $(a,t)$-deformed $F_K$ at radial limit, it is natural to consider a $t$-deformation of the $p$-th ADO polynomial.
The formal definition is given in \cite{Ekholm:2020lqy} as,
\begin{equation}\label{ADOdef}
   \operatorname{ADO}_K(p; x, t) = \Delta_K(x^p, - (-t)^p) \lim_{q \to \z_p} F_K(x, -t^{-1}q^2, q, t).
\end{equation}
In this subsection, we derive closed-form expressions of $t$-deformed ADO polynomials of the double twist knots. As in Appendix \ref{Appendix1}, it turns out that the closed formulae \eqref{Kmn} and \eqref{km+1/2} of cyclotomic type are suitable to compute the ADO polynomial up on the limit $q\rightarrow \z_p = e^{\frac{2\pi i}{p}}$ in which an infinite sum truncates to a finite one \cite{Anna:2020non}. (See also \cite{Banerjee:2020dqq} for a similar topic.)

First, we will obtain a closed-form expression of $t$-deformed Alexander polynomials.
Although $\Delta_K(x, t)$ can be easily calculated from the superpolynomial by using \eqref{alex-super}, we compute it from \eqref{Kmn} and \eqref{km+1/2} via the substitution \eqref{tAlexander}.
Notice that both the formulae share a common term $(- a t / q; q)_k$, where $k$ runs from $0$ to $\infty$. Upon specialization $a=-t^{-1}$, it becomes $(q^{-1}; q)_k$ and vanishes for $k > 1$. Therefore, the infinite summation becomes a rather simple finite one. Subsequently, they are given by
\begin{align}\label{T1}
   \Delta_{K_{m,n}}(x, t) &= -\frac{1}{t} + \left(t + \frac{1}{t} - t x - \frac{1}{t x}\right) S_m(-t) S_n(-t)~,\cr
   \Delta_{K_{m+\frac12,-n}}(x,t) &= t^n [n]_{tx}-t^{n-1}[n+1]_{tx} - t^n [n]_{tx}
   \left(t+\frac{1}{t} - t x -\frac{1}{t x}\right)S_m(-t)~,
\end{align}
where $S_l(x) = \sum\limits_{i = 0}^{l-1} x^i$, and the definition of quantum number $[n]_{tx}$ is given in \eqref{def:qno}.
We present our formulae this way, so that
the  Weyl symmetry \cite{Ekholm:2020lqy} of the $t$-deformed Alexander polynomial  becomes manifest,
\begin{equation}
   \Delta_K(x^{-1},t)=\Delta_K(t^{-2}x,t)~.
\end{equation}

After detailed derivation shown in Appendix \ref{Appendix1}, we obtain closed-from expressions of the $t$-deformed ADO polynomials for double twist knots $K_{m,n}$ and $K_{m+\frac12,-n}$,
\begin{align}
\operatorname{ADO}_{K_{m,n}}(p; x, t) &= (- t x)^{1 - p} \sum_{k = 0}^{p - 1} \lim_{q \to \z_p}g_{K_{m,n}}^{(k)} (x, -t^{-1}q^2, q, t), \cr
   \operatorname{ADO}_{K_{m+\frac12,-n}}(p; x, t) &= (- t x^n)^{1 - p} \sum_{k = 0}^{p - 1}\lim_{q \to \z_p} g_{K_{m + \frac12,-n}}^{(k)} (x, -t^{-1}q^2, q, t).
\end{align}
Note that there is no dependence on $t$-deformed Alexander polynomial because the radial limit of $(a,t)$-deformed $F_K$ is proportional to the inverse of the factor $\Delta_K(x^p, -(-t)^p)$, which cancels the same factor in \eqref{ADOdef}.
Especially, we can analytically derive that these formulae give rise to
\begin{align}
   \operatorname{ADO}_{K}(p = 1; x, t) & = 1, \cr
   \operatorname{ADO}_{K}(p = 2; x, t) & = \Delta_K (x, t),
\end{align}
which implies that the $t$-deformed ADO polynomial is a generalization of $t$-deformed Alexander polynomial.

In Table \ref{table3}, \ref{table4}, \ref{table5}, we summarize the results of $t$-deformed ADO polynomials for the trefoil, $5_2$ knot and $7_2$ knot.
\linespread{1.5}
\begin{table}[t]
\centering
\begin{tabular}[h]{lc}
$p$ & $\operatorname{ADO}_{3_1}(p; x, t)$\\
\hline \hline 1 & 1 \\
2 & $\left(-t x\right) + t + \left(-t x\right)^{-1}$ \\
3 & $\zeta_{3}^2\left[\left(t \zeta_{3} x\right)^{2}+t\left(t \zeta_{3} x\right)+\left(t^2-\zeta_{3}^{-1}\right)+t\left(t \zeta_{3} x\right)^{-1}+\left(t \zeta_{3} x\right)^{-2}\right]$ \\
\hline \hline
\end{tabular}
\caption{$t$-deformed ADO polynomials for the left-handed trefoil}
\label{table3}
\end{table}

\begin{table}[t]
\centering
\begin{tabular}[h]{lc}
$p$ & $\operatorname{ADO}_{5_2}(p; x, t)$\\
\hline \hline 1 & 1 \\
2 & $(1-t)(-t x) -1 + t -t^2 +(1-t) (-t x)^{-1}$ \\
3 & $\zeta_{3}^2\big[\left(1+t \zeta_{3} +t^2 \right)\left(t \zeta_{3} x\right)^{2}+\left(-1+2t+t^2 \zeta_{3}+t^3\right)\left(t \zeta_{3} x\right)+2+\zeta_{3}+(-1+\zeta_{3})t$\\
~ & $+2t^2+t^3 \zeta_{3}+t^4+\left(-1+2t+t^2 \zeta_{3}+t^3\right)\left(t \zeta_{3} x\right)^{-1}+\left(1+t \zeta_{3} +t^2 \right)\left(t \zeta_{3} x\right)^{-2}\big]$\\
\hline \hline
\end{tabular}
\caption{$t$-deformed ADO polynomials for the $5_2$ knot}
\label{table4}
\end{table}

\begin{table}[t]
\centering
\begin{tabular}[h]{lc}
$p$ & $\operatorname{ADO}_{7_2}(p; x, t)$\\
\hline \hline 1 & 1 \\
2 & $(1 - t + t^2)(- t x) - 1 + 2 t - t^2 + t^3 + (1 - t + t^2) (- t x)^{-1}$ \\
3 & $\zeta_{3}^2\big[\left(1 + t \zeta_{3} + t^3 \zeta_{3} + t^4 \right)\left(t \zeta_{3} x\right)^{2}
+ \left(-1 + (1 - \zeta_{3})t + 2 \zeta_{3} t^2 + t^3 + \zeta_{3} t^4 +t^5 \right)\left(t \zeta_{3} x\right)$\\
~ & $+ 2 + \zeta_{3} + (-1 + 2 \zeta_{3}) t + (1 - \zeta_{3}) t^2 + 2\zeta_{3} t^3 + t^4 + \zeta_{3} t^5 + t^6$\\
~ & $+\left(-1 + (1 - \zeta_{3})t + 2 \zeta_{3} t^2 + t^3 + \zeta_{3} t^4 +t^5 \right)\left(t \zeta_{3} x\right)^{-1}+\left(1 + t \zeta_{3} + t^3 \zeta_{3} + t^4 \right)\left(t \zeta_{3} x\right)^{-2}$ \big]\\
\hline \hline
\end{tabular}
\caption{$t$-deformed ADO polynomials for the $7_2$ knot}
\label{table5}
\end{table}

More generally, the higher rank $t$-deformed ADO is  defined as \cite{Ekholm:2020lqy}
\begin{equation}\label{eq:higher-rank-ADO}
   \operatorname{ADO}_K^{\SU(N)}(p; x, t) = \Delta_K(x^p, - (-t)^p)^{N-1} \lim_{q \to \z_p} F_K(x, -t^{-1} q^N, q, t).
\end{equation}
It is also feasible to derive the closed-form expression for the $t$-deformed ADO$^{\SU(N)}_K(p;x,t)$ from (\ref{Kmn}) and (\ref{km+1/2})
\begin{align}
\operatorname{ADO}_{K_{m,n}}^{\SU(N)} (p; x, t)
=& ~ \frac{1}{p}\sum_{l = 0}^{p - 1}(-t x)^{l + 1 - p} \Delta_{K_{m,n}} (x^p, -(-t)^p)^{\frac{(N-2)(p - 1) + l}{p}} S_p(\z_p^{N-l-2})\cr&\times\sum_{k=0}^{p - 1} \lim_{q\rightarrow\zeta_p}g^{(k)}_{K_{m,n}}(x,-t^{-1}q^N,q,t), \cr
\operatorname{ADO}_{K_{m+\frac12,-n}}^{\SU(N)} (p; x, t) = &~
\frac{1}{p} \sum_{l=0}^{p - 1}(-t x^n)^{l + 1 - p} \Delta_{K_{m+\frac12,-n}}(x^p, -(-t)^p)^{\frac{(N-2)(p - 1)+l}{p}}\cr
& ~ \times S_p(\z_p^{N - l -2})
\sum_{k=0}^{p - 1}\lim_{q\rightarrow\zeta_p} g^{(k)}_{K_{m+\frac{1}{2},-n}}(x,-t^{-1}q^N,q,t).
\end{align}
The detailed derivation is also shown in the Appendix \ref{Appendix1}. Especially, we can analytically verify that they become
\begin{align}
   &\operatorname{ADO}_K^{\SU(N)}(1;x,t)=1,\cr
   &\mathrm{ADO}_{K}^{\SU(N)}(2 ; x, t)=\frac{1}{2}\sum_{l,j=0}^{1}(-1)^{j(N-l)}\left[\Delta_K(x^2,-(-t)^2)\right]^{\frac{l}{2}+\frac{N}{2}-1}\left[\Delta_{K}(x,t)\right]^{1-l}~,
\end{align}
and they obey the recursion relation  \cite{Ekholm:2020lqy}
\begin{align}
   \operatorname{ADO}^{\SU(N+p)}(p;x,t)=\Delta_K(x^p,-(-t)^p)^{p-1}\operatorname{ADO}^{\SU(N)}(p;x,t)~.
\end{align}

Note that colored Jones polynomials of cyclotomic type \cite{Habiro:2008,lovejoy2017colored,lovejoy2019colored} at a root of unity $q=\zeta_p$ provide the corresponding ADO invariants \cite{Anna:2020non} even if a closed-form expression of $F_K$ is not known. Therefore, one can evaluate the ADO invariants from colored Jones polynomials in a similar fashion for arbitrary double twist knots such as $K_{m,n}$ with $m,-n\in \bZ_+$ \cite{Anna:2020non}.

   \section{Conclusion and discussion}

     In this paper, we presented a simple rule of grading change that would allow us to obtain $[r]$-colored quadruply-graded Kauffman homology from $[r^2]$-colored quadruply-graded HOMFLY-PT homology for thin knots. We check the grading change by consistency with differentials and symmetries for colored Kauffman homology of double twist knots. We also find from the universal differential that the super-$A$-polynomials of SO-types contains both $a$-deformed $A$-polynomials of SO and SU-type at $t=-1$.
With the natural quantization of $\SO(N)(N \geq 4)$ $6j$-symbols for symmetric representations given by Ali\v{s}auskas, we calculate $\SO(N)$ fusion matrices for $R= ~$\yng(1) , \yng(2) and compute Kauffman polynomials for the corresponding colors. We check the validity of the quantization from representation theory and Kauffman polynomials.
     We  also conjecture a closed-form expression of the $(a,t)$-deformed $F_K$ of the double twist knots $K_{m,n}$ and $K_{m+\frac{1}{2},-n}$ from the corresponding superpolynomial. Using the $(a,t)$-deformed $F_K$ we proposed, we also gave closed-form expressions of the $t$-deformed Alexander and ADO polynomial.

      However, there are still immediate questions that need to be solved in the future. First, the expression \eqref{quantum6j} is rather involved and there must be a similar formula for it. It is also desirable to find a closed-form expression of $6j$-symbols when fusions $R_{12}$ and $R_{23}$ are non-symmetric. Moreover, it has been found that \cite{Liu:2018jhs} the $6j$-symbol  in $\text{AdS}$ is the Lorentzian inversion of a crossing-symmetric tree-level exchange amplitude, and the one-loop vertex correction in $\phi^3$-theory in $\text{AdS}_{d+1}$ is given by a spectral integral over the $6j$-symbol for $\SO(d+1,1)$.  It would be interesting that the results in \S\ref{section:6j-symbols} would bring some inspiration to the calculations of these $6j$-symbols in the $\text{AdS}$ background.

So far the study of the volume conjecture involves only the large color limit of symmetric representations. It is very important to formulate volume conjectures with arbitrary colors, relating to the moduli space of flat $\SL(N,\C)$ connections over the knot complement. Since colored knot homology is endowed with many colored differentials, we expect a behavior similar to Conjecture \ref{A-SO-SU} in volume conjectures of higher ranks.

We have obtained the $(a,t)$-deformed $F_K$ for the two classes of double twist knots from the corresponding superpolynomials. However, It is desirable to develop a general method to obtain a closed-form expression of $F_K$ for arbitrary knots.

   \acknowledgments
   We would like to thank Chen Yang for collaboration at the initial stage of the project.
   S.N. is indebted to Bruno le Floch for collaboration and discussion on $6j$-symbols, and he also thanks Ryo Suzuki for identifying the reference \cite{Alisauskas_2002} about $\SO(N)$ $6j$-symbols.
   We also would like to thank Sunghyuk Park for identifying the relationship between the $t$-deformed ADO polynomials of $3_1$ and $^*3_1$. This work was supported by the National Science Foundation of China under Grant No. NSFC PHY-1748958.

\appendix
   \section{Derivation of the ADO polynomials}\label{Appendix1}
   In this Appendix, we derive the $t$-deformed $p$-th ADO polynomial \eqref{ADOdef}
   and its higher rank generalization \eqref{eq:higher-rank-ADO} from $F_K(x,a,q,t)$
   \eqref{Kmn} and \eqref{km+1/2}.
   We use $F_K$ as a short for,
   $$\lim_{q \to \zeta_p} F_K(x, a = -t^{-1} q^N, q, t)~,$$
   and similarly for other functions discussed in this section.
   We use $q$ and $\z_p$ interchangeably.
   When an integer $k$ is displayed as $k = k_1 p + k_0$, it is implied that
   $k_1, k_0 \in \mathbb{N}$, and $0 \le k_0 \le p - 1$.

   Our closed formulae are built from $q$-binomials and $q$-Pochhammer symbols and we will first discuss their behavior when $q$ goes to roots of unity.
   The $q$-Lucas theorem \cite{olive1965generalized} states that,
   \be \label{ap:qBino}
    {l \brack k}_{q = \zeta_p} = {l_1 \choose k_1} {l_0 \brack k_0}_{q = \zeta_p}~,
   \ee
   where $l = l_1 p+ l_0,~k = k_1p + k_0$.
   For $q$-Pochhammers, we have
   \be \label{ap:qPoch}
   (a;q)_k = (1 - a^p)^{k_1} (a;q)_{k_0}~,
   \ee
   where $q = \zeta_p$, $k = k_1 p + k_0$. In the following discussion, We will break $F_K(x,a,q,t)$ into parts. As we will see later, these components enjoy similar properties. It turns out that $F_K$ can be written as a product of a infinite summation over $k_1$ and a finite one over $k_0$, where $k = k_1p +k_0$ and $q = \zeta_p$, and the infinite summation over $k_1$ can be repackaged by the generalized binomial theorem,
   \be
   \frac{1}{(1-z)^n} = \sum_{i = 0}^\infty {n + i -1 \choose i} z^i~.
   \ee
   As a result, an ADO polynomial can be expressed as a finite summation of over $k_0$.

   \subsection{ADO polynomials of double twist knot $K_{m,n}~(m,n\in \mathbb{Z}_+)$}
   For double twist knots $K_{m,n}~(m,n \in \mathbb{Z}_+)$, we first analyze $(a,t)$-deformed $F_K$ at radial limit $q = \zeta_p$. Each component of \eqref{Kmn} behaves as follows.
   \begin{itemize}
       \item The twist factor $\operatorname{tw}^{(l)}_m(a,q,t)$ \eqref{eq:tw} now becomes
       \be \label{ap:tw}
       \operatorname{tw}_m^{(l)} =
       \sum_{0\leq b_1 \leq \cdots \leq b_{m-1} \leq b_m = l}
       \prod_{i=1}^{m-1}(-q^N)^{b_i} q^{b_i (b_i-1)} t^{ b_i}
       {b_{i+1} \brack b_{i}}_q~.
       \ee
       We decompose the summation variables as, $b_{i} = \alpha_i p + \beta_i$, and $l = l_1 p + l_0$.
       Using the $q$-Lucas theorem, \eqref{ap:tw} can be written as a product of two summations, one over $\alpha_i$'s and another one over $\beta_i$'s.
       The summation over $\beta_i$'s would give $\mathrm{tw}_m^{(l_0)}$.
       Performing the summation over $\alpha_i$'s, we obtain
       \be \label{twist-factor}
       \mathrm{tw}_m^{(l)} = S_m^{l_1}((-t)^p)\,\mathrm{tw}_m^{(l_0)}~,
       \ee
       where $S_m^{l_1}(x) := \left(S_m(x)\right)^{l_1} := \left(\sum\limits_{i = 0}^{m-1}x^i\right)^{l_1}$.

       \item The twist factor $\operatorname{Tw}_{K_{m,n}}^{(j)}(a,q,t)$ \eqref{twist2} becomes
       \bea
       \operatorname{Tw}_{K_{m,n}}^{(j)} = \sum_{l = 0}^j
       (-1)^l q^{\frac{1}{2} l (l + 1) - j l} \operatorname{tw}_m^{(l)}
       \operatorname{tw}_n^{(l)} {j \brack l}_q~.
       \eea
       We write the summation variables as $j = j_1 p + j_0$, and $l = l_1 p + l_0$. Because of the fact that $q^p = 1$, the $q$-Lucas theorem and (\ref{twist-factor}), we obtain,
       \be \label{ap:Tw}
       \operatorname{Tw}_{K_{m,n}}^{(j)} = \left(1-S_m\left(\left(-t\right)^p\right)S_n\left(\left(-t\right)^p\right)\right)^{j_1} \operatorname{Tw}_{K_{m,n}}^{(j_0)}~.
       \ee

       \item Now $g^{(k)}_{K_{m,n}}(x, a, q, t)$ in \eqref{eq:g}  becomes
       \be \label{Appendix:g}
       g^{(k)}_{K_{m,n}} = \sum_{j = 0}^k (x;q^{-1})_k
       (x t^2 q^N; q)_j (x t^2)^{k-j} q^{k-Nj}
       {N-2+k \brack k}_q {k \brack j}_q
       \operatorname{Tw}_{K_{m,n}}^{(j)}~,
       \ee
       where we have used the fact that
       \be
        \frac{(q^{N-1};q)_k}{(q;q)_k} = {N-2+k \brack k}_q~.
       \ee
       We decompose the variables as $j = j_1 p + j_0$, $k = k_1 p + k_0$ and $N-2 = A_1 p + A_0$.
       Plugging \eqref{ap:qBino}, \eqref{ap:qPoch} and \eqref{ap:Tw} into \eqref{Appendix:g}, we have
       \be
       g_{K_{m,n}}^{(k)} ={A_1 + k_1 \choose k_1}
       \left(1 - x^p\right)^{k_1}
       \left[1- \left(1 - x^p t^{2p} \right)S_m\left((-t)^p\right)S_n\left((-t)^p\right)\right]^{k_1} g^{(k_0)}_{K_{m,n}}~.
       \ee

       \item $F_{K_{m,n}} (x, a, q, t)$ becomes
       \be
       F_{K_{m,n}} = (- t x)^{N-1} \sum_{k = 0}^\infty g_{K_{m,n}}^{(k)}~.
       \ee
       Given $k = k_1 p + k_0$, $N - 2 = A_1 p + A_0$, we have
       \bea
       F_{K_{m,n}} =&~ (- t x)^{N - 1}
       \sum_{k_0 = 0}^{p-1} g_{K_{m,n}}^{(k_0)} \cr
       &\times \sum_{k_1 = 0}^\infty {A_1 + k_1 \choose k_1}
       \left(1 - x^p\right)^{k_1}
       \left[1- \left(1 - x^p t^{2p}\right)S_m\left((-t)^p\right)S_n\left((-t)^p\right)\right]^{k_1}~.
       \eea
       Using the generalized binomial theorem, we obtain
       \be
       F_{K_{m,n}} = \frac{(- t x)^{N-1} \sum\limits_{k = 0}^{p-1}g^{(k)}_{K_{m,n}}}{\left[x^p + \left( 1 - x^p\right)
       \left( 1- x^pt^{2p}\right) S_m\left(\left(-t\right)^p\right) S_n\left(\left(-t\right)^p\right)\right]^{A_1 + 1}}.
       \ee
       Recall the closed formulae of $t$-deformed Alexander polynomials \eqref{T1}, we can write
       \be
       F_{K_{m,n}} = \frac{(-t x)^{A_0 + 1 - p}}{\Delta_{K_{m,n}}(x^p, -(-t)^p)^{A_1 + 1}} \sum_{k = 0}^{p-1} g^{(k)}_{K_{m,n}}~.
       \ee
    \end{itemize}
    Before we jump to the ADO polynomials, let us first examine the properties of $(a,t)$-deformed $F_K$.
       When $p=1$ which leads to $A_1 = N-2$, $A_0 = 0$, we have
       \be
       \lim_{q \to 1} F_{K_{m,n}}(x, -t^{-1}q^N, q, t) = \frac{g^{(0)}_{K_{m,n}}}{\Delta_{K_{m,n}}(x,t)^{N-1}}
       = \frac{1}{\Delta_{K_{m,n}}(x,t)^{N-1}}~,
       \ee
       in accordance with \eqref{property3}.
       Finally, for the $t$-deformed $\operatorname{ADO}$ polynomials, we have
       \bea
       \operatorname{ADO}_{K_{m,n}}^{\SU(N)}(p; x, t) &= \Delta_{K_{m,n}}(x^p, -(-t)^p)^{N-1} F_{K_{m,n}} \cr &=
       (-tx)^{A_0 + 1 - p} \Delta_{K_{m,n}}(x^p, - (-t)^p)^{A_1(p-1) + A_0}
       \sum_{k=0}^{p-1} g_{K_{m,n}}^{(k)}~,
       \eea
       where $N-2 = A_1 p + A_0$.

Now let us consider some simple cases. If $p=1$, then $A_0 = 0$, we have
\be
    \operatorname{ADO}_{K_{m,n}}^{\SU(N)}(p=1; x,t) = g^{(0)}_{K_{m,n}} = 1~.
\ee
For $N = 2$ and $p = 2$, then $A_0 = A_1 = 0$, we have
\be
    \operatorname{ADO}_{K_{m,n}}^{\SU(2)}(p=2; x,t) = (-t x)^{-1} \left(g^{(0)}_{K_{m,n}} + g^{(1)}_{K_{m,n}}\right) = \Delta_{K_{m,n}}(x,t)~.
\ee

\subsection{ADO polynomials of double twist knots $K_{m + \frac12, -n}~(m,n \in \mathbb{Z}_+)$}
Following the same procedure in the last subsection, we decompose the closed-form expression of $F_{K_{m+\frac12,-n}}  (x,a,q,t)$ \eqref{km+1/2} into parts:
\begin{itemize}
   \item The twist factor $\tw_n^{(k)}(x, q, t)$ \eqref{eq:TW} now becomes
   \be
   \tw_n^{(k)} (x, q, t) = \sum_{0 = b_0 \le b_1 \le \cdots \le b_{n-1} \le b_n = k}
   \prod_{i=1}^{n-1} (x^{2 b_i} q^{- b_i (b_{i+1}-1)}t^{2b_i})
   {b_{i + 1} \brack b_i}_q~.
   \ee
   We write $b_i = \a_i p + \b_i$, $k = k_1 p + k_0$, and we can obtain
   \be
   \tw_n^{(k)}
    = S^{k_1}_n(x^{2p} t^{2p}) \,\tw_n^{(k_0)}~.
   \ee

   \item Now  $g^{(k)}_{K_{m+\frac12,-n}}$ in \eqref{eq:g2} becomes
   \bea
   g^{(k)}_{K_{m+\frac12,-n}} = \sum_{j = 0}^k (x; q^{-1})_k
    (x t^2 q^N; q)_j x^{k-j} {N-2+k \brack k}_q
   {k \brack j}_q q^{\frac12j(j-1)+k} t^{-j + 2k} \operatorname{tw}_m^{(j)} \tw_n^{(k)}~.
   \eea
   We write $k = k_1 p + k_0$, $j = j_1 p + j_0$, and $N - 2 = A_1 p + A_0$, and obtain
   \bea
   g^{(k)}_{K_{m+\frac12, -n}}
   =&~ \left\{
   \left(1- x^p\right)\left(-t\right)^p
   \left[
   \left(-x t\right)^p + \left(x^p t^{2p} - 1\right)
   S_m \left(\left(- t\right)^p\right)
   \right]
   S_n\left(x^{2p} t^{2p}\right)
   \right\}^{k_1}\cr
   &\times {A_1 + k_1 \choose k_1}
   g_{K_{m+\frac12,-n}}^{(k_0)}~.
   \eea

   \item $F_{K_{m+\frac12,-n}}(x,a,q,t)$ now becomes
   \be
   F_{K_{m+\frac12,-n}} = (-t x^n)^{N-1} \sum_{k=0}^{\infty} g_{K_{m+\frac12,-n}}^{(k)}~.
   \ee
   We write $k = k_1 p + k_0$, $N-2 = A_1 p + A_0$. Again, with the help of the generalized binomial theorem, we can get rid of the infinite summation over $k_1$ and obtain
   \be
   F_{K_{m+\frac12,-n}} = \frac{(-t x^n)^{A_0 + 1 - p}}{\Delta_{K_{m+\frac12,-n}}(x^p, -(-t)^p)^{A_1 + 1}} \sum_{k_0 = 0}^{p - 1} g^{(k_0)}_{K_{m+\frac12,-n}}~,
   \ee
   where the $t$-deformed Alexander polynomial is given in \eqref{T1}.
\end{itemize}
When $p = 1$, then $A_1 = N - 2$, $A_0 = 0$, we have
\be
\lim_{q\to 1}F_{K_{m+\frac12, -n}} (x,-t^{-1}q^N,q,t) = \frac{g^{(0)}_{K_{m+\frac12,-n}}}{\Delta_{K_{m+\frac12,-n}}(x,t)^{N-1}}
=\frac{1}{\Delta_{K_{m+\frac12,-n}}(x,t)^{N-1}}~,
\ee
in accordance with \eqref{property3}.
Finally, for the $t$-deformed $\operatorname{ADO}$ polynomials, we have
\bea
\operatorname{ADO}_{K_{m+\frac12,-n}}^{\SU(N)}(p; x, t) &= \Delta_{K_{m+\frac12,-n}}(x^p, -(-t)^p)^{N-1} F_{K_{m + \frac12,-n}} \cr &=
(-tx^n)^{A_0 + 1 - p} \Delta_{K_{m + \frac12,-n}}(x^p, - (-t)^p)^{A_1(p-1) + A_0}
\sum_{k=0}^{p-1} g_{K_{m+\frac12,-n}}^{(k)}~,
\eea
where $N-2 = A_1 p + A_0$.

Now we consider some simple cases. When $p = 1$, $A_0 = 0$, we have
\be
\operatorname{ADO}_{K_{m+\frac12, -n}}^{\SU(N)}(p = 1; x, t) = g^{(0)}_{K_{m+\frac12,-n}} =1~.
\ee
For $N = 2$, $p = 2$, we have
\be
\operatorname{ADO}_{K_{m+\frac12,-n}}(p = 2; x, t) = (-t x^n)^{-1}
(g^{(0)}_{K_{m+\frac12,-n}} + g^{(1)}_{K_{m+\frac12,-n}}) = \Delta_{K_{m+\frac12, -n}} (x, t)~.
\ee

\subsection{Final formulae}
In conclusion, for gauge group $\SU(N)$, the $t$-deformed $p$-th ADO
polynomials are given by
\begin{align}
    &\operatorname{ADO}_{K_{m,n}}^{\SU(N)} (p; x, t)
    = (-t x)^{A_0 + 1 - p} \Delta_{K_{m,n}}(x^p, -(-t)^p)^{A_1 (p - 1) + A_0}
    \sum_{k=0}^{p - 1} g^{(k)}_{K_{m,n}}, \cr
    &\operatorname{ADO}_{K_{m+\frac12,-n}}^{\SU(N)} (p; x, t) = (-t x^n)^{A_0 + 1 - p} \Delta_{K_{m+\frac12,-n}}(x^p, -(-t)^p)^{A_1 (p - 1)+ A_0}
    \sum_{k=0}^{p - 1} g^{(k)}_{K_{m + \frac12,-n}}, \nonumber
\end{align}
where $N-2 = A_1 p + A_0$, $A_1, A_0 \in \mathbb{N}$ and $0\le A_0 \le p-1$. It is easily seen that the recursion relation conjectured in \cite{Ekholm:2020lqy} holds,
\be
    \operatorname{ADO}^{\SU(N+p)}_{K}(p; x, t) = \Delta_K(x^p, -(-t)^p)^{p-1} \operatorname{ADO}^{\SU(N)}_{K}(p; x, t)~.
\ee

   Note that $\frac{1}{p}S_m(\z_p^n) = 1$, only when $m | n$. Otherwise it is zero. Therefore, the $t$-deformed $p$-th ADO polynomials, can also be written as,
   \bea
   \operatorname{ADO}_{K_{m,n}}^{\SU(N)} (p; x, t)
   =& ~ \frac{1}{p}\sum_{l = 0}^{p - 1}(-t x)^{l + 1 - p} \Delta_{K_{m,n}}(x^p, -(-t)^p)^{\frac{(N-2)(p - 1) + l}{p}}
   S_p(\z_p^{N-l-2})
   \sum_{k=0}^{p - 1} g^{(k)}_{K_{m,n}} \cr
   \operatorname{ADO}_{K_{m+\frac12,-n}}^{\SU(N)} (p; x, t) = &~
   \frac{1}{p} \sum_{l=0}^{p - 1}(-t x^n)^{l + 1 - p} \Delta_{K_{m+\frac{1}{2},-n}}(x^p, -(-t)^p)^{\frac{(N-2)(p - 1)+l}{p}}\cr
   & ~ \times S_p(\z_p^{N - l -2})
   \sum_{k=0}^{p - 1} g^{(k)}_{K_{m+\frac{1}{2},-n}}.
   \eea
   Although we write them as summations over $l$, there is only one non-vanishing term, which corresponds to that $l$ is the remainder of $(N-2)/p$.

\bibliography{CS}{}
\bibliographystyle{JHEP}

\end{document}